\pdfoutput=1
\documentclass[11pt,a4paper]{article}
\usepackage{jheppub}

\RequirePackage[utf8]{inputenc}

\usepackage{amsfonts}
\usepackage{amssymb}
\usepackage{comment}
\usepackage{graphicx}
\usepackage{amsmath}
\usepackage{tabu}
\usepackage{dsfont}
\usepackage{float}
\usepackage{amsthm}
\usepackage{slashed}
\usepackage{setspace}
\usepackage[labelformat=simple]{subcaption}

\usepackage[boxsize=0.5em,aligntableaux=center]{ytableau}

\newlength{\PicScale}
\newcommand{\sG}{\mathcal{O}}
\newcommand{\cF}{\mathcal{F}}
\newcommand{\cV}{\mathcal{V}}
\newcommand{\cD}{\mathcal{D}}
\newcommand{\cC}{\mathcal{C}}
\newcommand{\bea}{\begin{eqnarray}}
\newcommand{\eea}{\end{eqnarray}}
\newcommand{\be}{\begin{equation}}
\newcommand{\ee}{\end{equation}}
\newcommand{\ba}{\begin{align}}
\newcommand{\ea}{\end{align}}
\newcommand{\comments}[1]{}
\def\nn{\nonumber}

\def\LVS{{\scriptscriptstyle LVS}}

\newcommand\vo{{\mathcal{V}}}

\newcommand{\mc}{\mathcal}

\newcommand{\beqa}{\begin{eqnarray}}
\newcommand{\eeqa}{\end{eqnarray}}

\newcommand{\Tr}[1]{\textmd{Tr}\left[#1\right]}

\def\Tr{\text{Tr}}

\newcommand{\nc}{\newcommand}
\nc{\beq}{\begin{equation}}
\nc{\eeq}{\end{equation}}
\def\IZ{\mathbb{Z}}
\def\ov{\overline}
\newcommand{\rd}[1]{\mathbf{#1}}
\newcommand{\bC}{\mathbb{C}}

\newcommand{\cI}{\mathcal{I}}

\newcommand{\cR}{\mathcal{R}}

\newcommand{\symm}{\ydiagram{2}}
\newcommand{\asymm}{\ydiagram{1,1}}
\newcommand{\fund}{\ydiagram{1}}

\newcommand{\adj}{\ensuremath{\text{Adj}}}

\newcommand{\Sp}{U\!Sp}

\newcommand{\SU}{SU}

\newcommand{\U}{U}

\allowdisplaybreaks[2]
\numberwithin{equation}{section}

\title{Global Orientifolded Quivers with Inflation}

\author[1,2,3]{Michele Cicoli,}
\author[4]{I\~naki Garc{\'i}a-Etxebarria,}
\author[5]{Christoph Mayrhofer,}
\author[3,6]{Fernando Quevedo,}
\author[3]{Pramod Shukla,}
\author[7,8,3]{Roberto Valandro}

\affiliation[1]{Dipartimento di Fisica e Astronomia, Universit\`a di Bologna, via Irnerio 46, 40126 Bologna, Italy}
\affiliation[2]{INFN, Sezione di Bologna, viale Berti Pichat 6/2, 40127 Bologna, Italy}
\affiliation[3]{ICTP, Strada Costiera 11, Trieste 34151, Italy}
\affiliation[4]{Max Planck Institute for Physics, F\"ohringer Ring 6, 80805 Munich, Germany}
\affiliation[5]{Arnold Sommerfeld Center for Theoretical Physics, Theresienstrae 37, 80333 M\"unchen, Germany}
\affiliation[6]{DAMTP, Centre for Mathematical Sciences, Wilberforce Road, Cambridge, CB3 0WA, UK.}
\affiliation[7]{Dipartimento di Fisica, Universit\`a di Trieste, Strada Costiera 11, 34151 Trieste, Italy}
\affiliation[8]{INFN, Sezione di Trieste, Via Valerio 2, 34127 Trieste, Italy}
\emailAdd{mcicoli@ictp.it} \emailAdd{inaki@mpp.mpg.de}\emailAdd{christoph.mayrhofer@lmu.de} \emailAdd{f.quevedo@damtp.cam.ac.uk}
\emailAdd{shukla.pramod@ictp.it}\emailAdd{roberto.valandro@ts.infn.it}

\abstract{We describe global embeddings of fractional D3 branes at orientifolded singularities in type IIB flux compactifications. We present an explicit Calabi-Yau example where the chiral visible sector lives on a local orientifolded quiver while non-perturbative effects, $\alpha'$ corrections and a T-brane hidden sector lead to full closed string moduli stabilisation in a de Sitter vacuum. The same model can also successfully give rise to inflation driven by a del Pezzo divisor. Our model represents the first explicit Calabi-Yau example featuring both an inflationary and a chiral visible sector.}


\begin{document}

\makeatletter
\let\old@fpheader\@fpheader
\renewcommand{\@fpheader}{\old@fpheader\hfill
DAMTP-2017-21

\hfill MPP-2017-124}
\makeatother

\maketitle

\bigskip

\section{Introduction}

Fractional BPS D-branes at Calabi-Yau (CY) singularities provide a powerful realisation of gauge theories in string theory. These configurations have been useful in the study of gauge/gravity dualities and as starting points for fully-fledged string compactifications with chiral matter. The  gauge theories corresponding to a collection of fractional branes are properly represented in terms of quiver diagrams which are two-dimensional graphs where nodes, representing the fractional branes, are joined by lines representing matter fields transforming in bi-fundamental representations of the corresponding gauge symmetries. 

In the absence of an orientifold action the graph is oriented with outgoing/ingoing arrows representing fundamental/anti-fundamental representations of the corresponding gauge group at the node. Orientifold involutions change the orientation of the diagram and the associated gauge theory with $U(N)$ groups are either identified with each other under the involution or projected down to $SO(N)/\Sp(N)$ if the involution acts non-trivially on the corresponding node. The orientation of the arrows is changed appropriately and the quiver is generically no longer oriented.

In order to have a complete string theory background that is phenomenologically viable, the quiver gauge theory has to be globally embedded in a consistent compactification. In type IIB the most interesting compactifications are Calabi-Yau threefolds with a proper orientifold involution that leads to $N=1$ supersymmetry. This generically implies the presence of O7 and/or O3-planes that carry RR charges. To cancel those charges, collections of D-branes have to be added at different locations in the compact space. Furthermore gauge fluxes of different antisymmetric tensors have to be turned on to satisfy consistency conditions. These fluxes induce non-trivial F- and D-terms in the effective 4D theory that help to stabilise the geometric moduli. Finally the presence of ED3-instantons gives rise to non-vanishing superpotentials that, together with fluxes and $\alpha'$ effects, can stabilise all moduli fields in a de Sitter vacuum.

The construction of explicit compactifications with quiver gauge theories on fractional D3 branes at singularities, is a very promising way to go for string phenomenology since it can lead to a chiral 4D visible sector. A consistent global embedding of them should be characterised by full moduli stabilisation in an almost Minkowski vacuum and interesting cosmological implications. A successful embedding of oriented quiver theories in compact Calabi-Yau orientifolds has already been presented in \cite{Cicoli:2012vw, Cicoli:2013mpa, Cicoli:2013zha, Cicoli:2013cha}. In this article we extend this approach by showing how to do that for orientifolded quivers. We will exemplify this class of string constructions by presenting a concrete model where a global embedding and moduli stabilisation are explicitly realised.

\begin{figure}
\centerline{\psfig{file=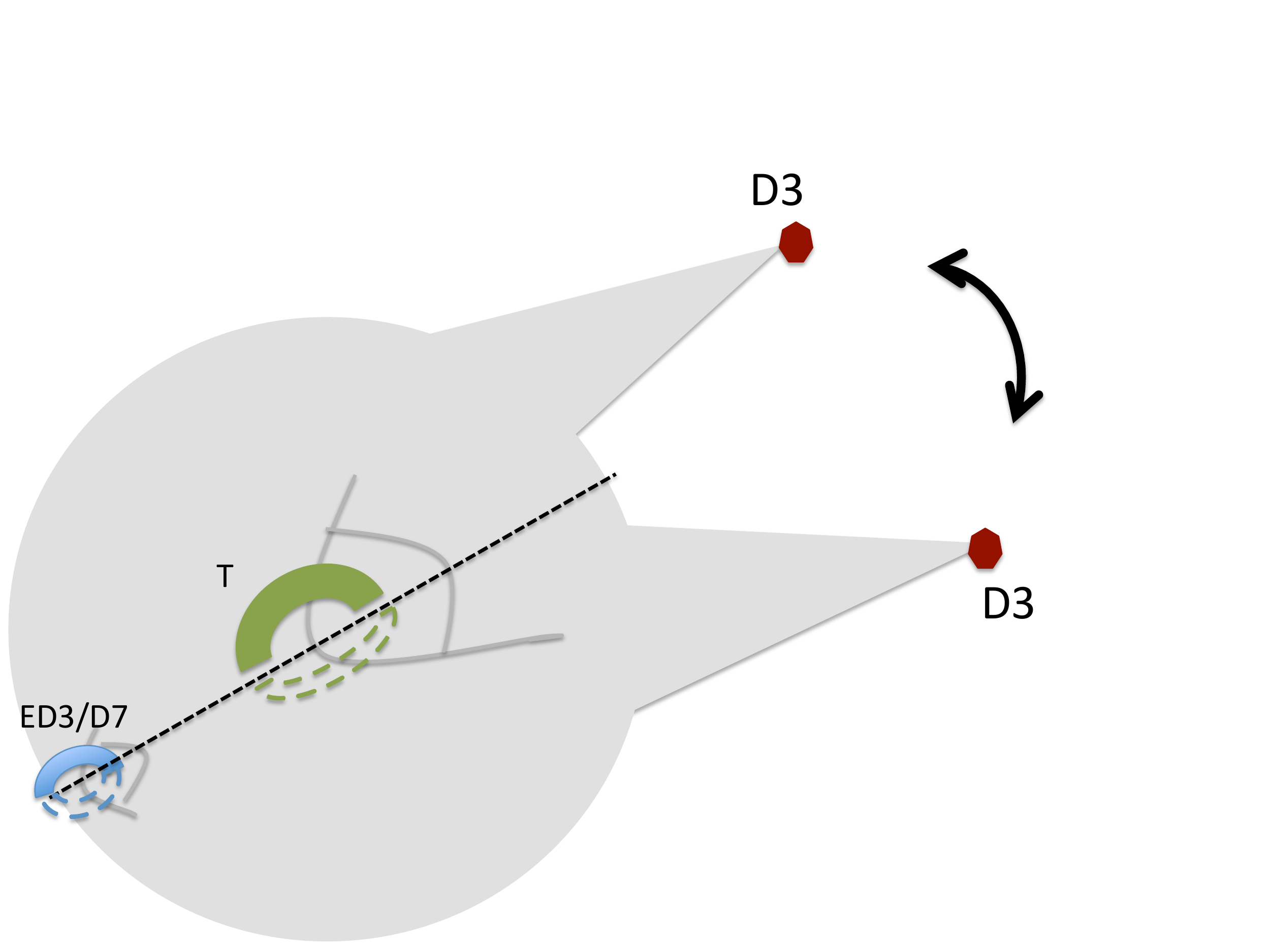,width=8.8cm}}
\caption{Global embedding of a local oriented quiver coming from fractional D3 branes at singularities. The action of the orientifold involution is represented by the dashed line. The involution exchanges two identical quivers. An additional del Pezzo divisor can support either an ED3-instanton or a D7 stack with gaugino condensation. Due to the presence of non-zero gauge fluxes, the large four-cycle tends to be wrapped by a hidden D7 stack (T-brane) which is responsible for a dS vacuum.}
\label{ra_fig1}
\end{figure}

Besides the natural motivation to fill the gap on this general class of string compactifications, orientifolded quivers have several interesting phenomenological properties:
\begin{itemize}
\item The minimal quiver extensions of the supersymmetric Standard Model consists of three or  four-node quivers which are unoriented, for some nodes there are only incoming or outgoing arrows. This cannot be obtained from standard oriented quivers (see for instance \cite{Wijnholt:2007vn, Berenstein:2006pk}).

\item Orientifolded quivers are more generic than oriented ones since, from the global embedding point of view, they do not require Calabi-Yau threefolds with two identical singularities mapped to each other under the orientifold involution.

\item Concrete local models of dP$_n$ quivers can, after Higgsing, give rise to semi-realistic extensions of the Standard Model without the need of flavour D7-branes.
\end{itemize}

\begin{figure}
\centerline{\psfig{file=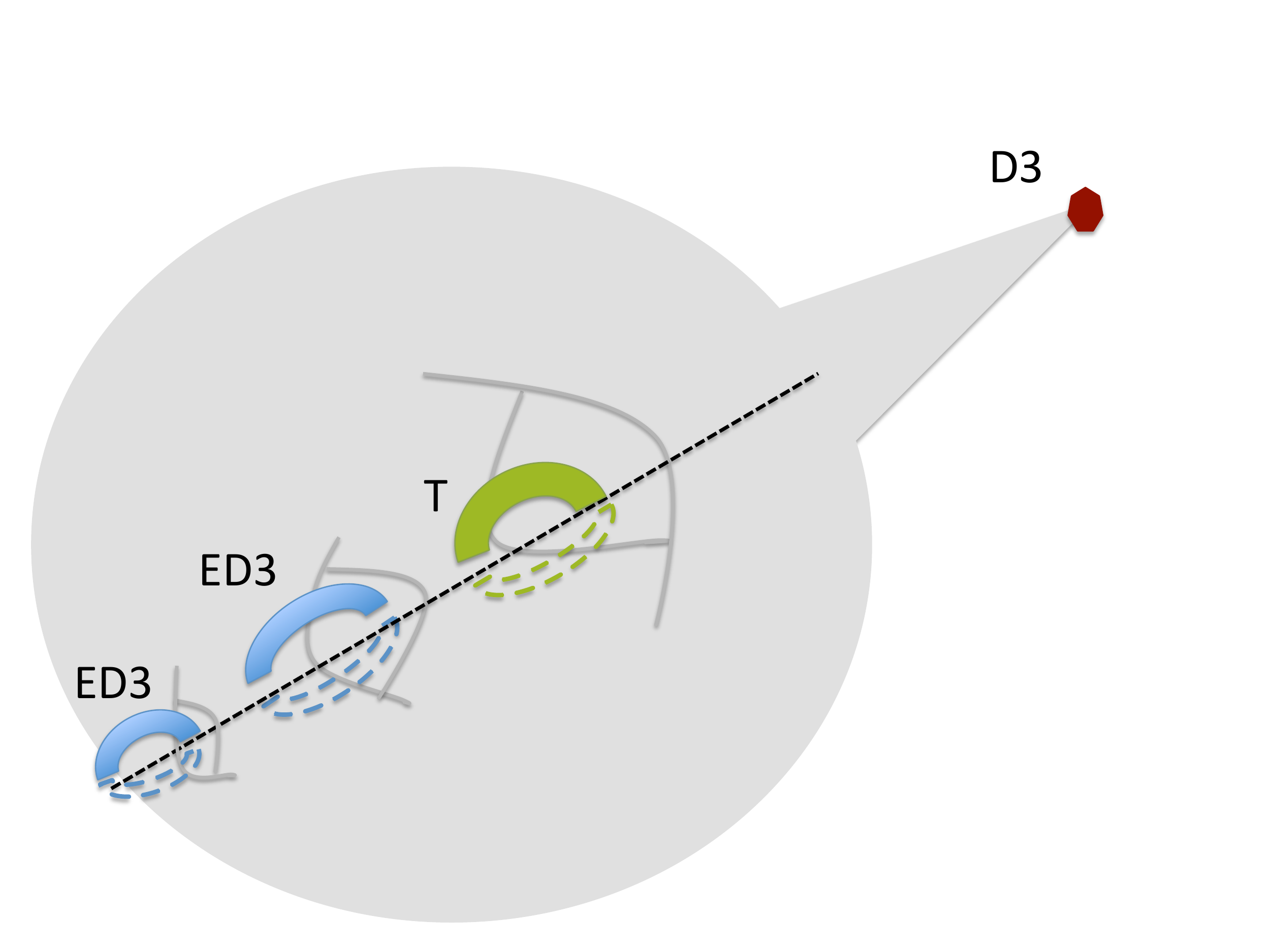,width=8.8cm}}
\caption{Global embedding of an orientifolded quiver. The action of the orientifold involution is represented by the dashed line. The two del Pezzo divisors in the geometric regime support ED3-instantons while the large four-cycle is wrapped by a hidden D7 stack (T-brane) which is responsible for a dS vacuum. Both the ED3-instantons and the D7 T-brane wrap invariant divisors.}
\label{ra_fig2}
\end{figure}

\begin{figure}
\centerline{\psfig{file=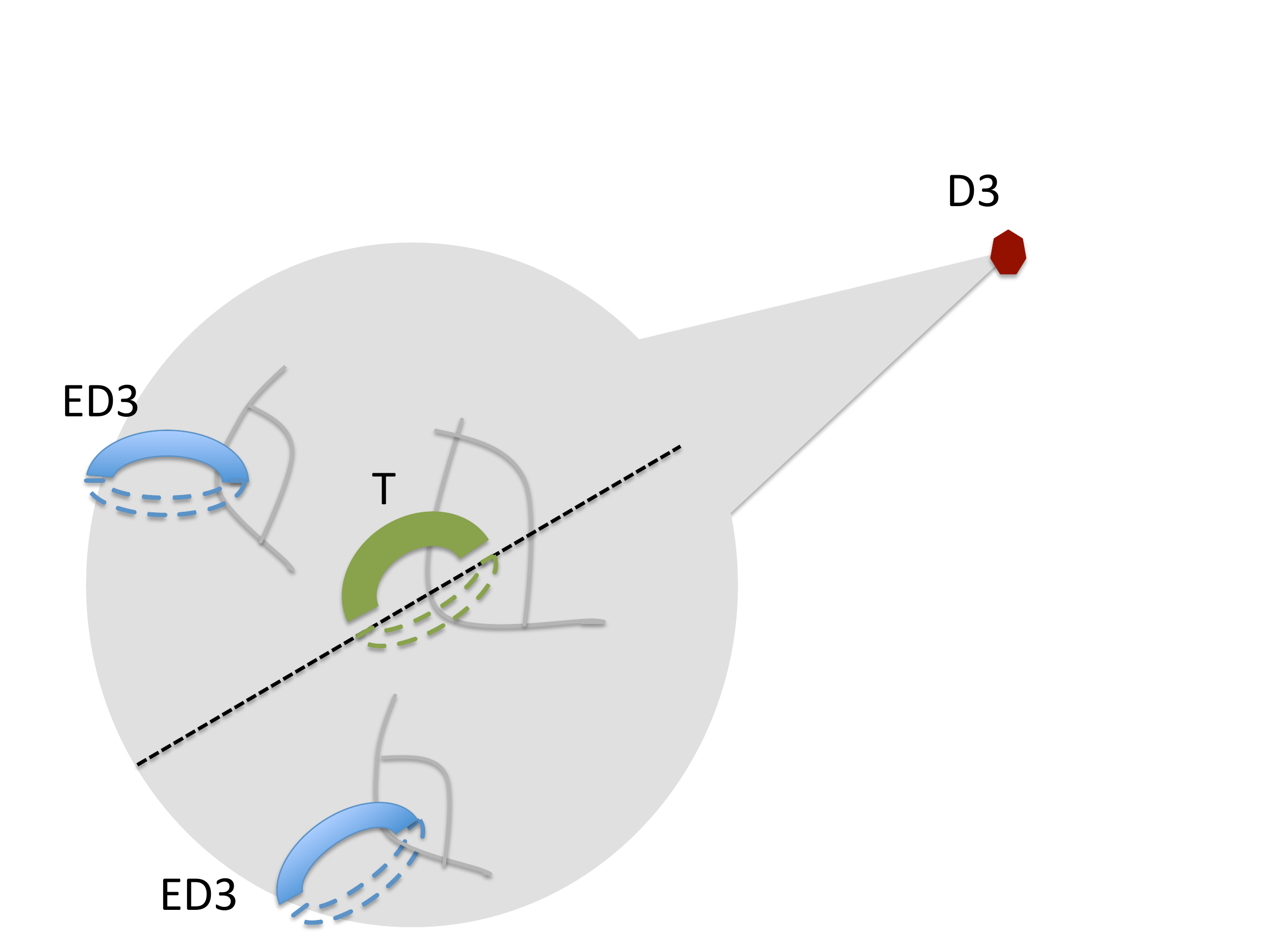,width=8.8cm}}
\caption{Global embedding of an orientifolded quiver. The action of the orientifold is represented by the dashed line. The two del Pezzo divisors in the geometric regime are wrapped by ED3-instantons which are exchanged under the involution, and so lead to a $U(1)$ instanton. The large (orientifold invariant) cycle is instead wrapped by a D7 T-brane stack that gives rise to a dS vacuum.}
\label{ra_fig3}
\end{figure}

A pictorial representation of the two classes of string compactifications for the case with four K\"ahler moduli is illustrated in Fig.~\ref{ra_fig1},  \ref{ra_fig2} and \ref{ra_fig3}. We focus on cases where the volume of the Calabi-Yau manifold admits a typical Swiss-cheese form: $\vo=\tau_1^{3/2}-\tau_2^{3/2}-\tau_3^{3/2}-\tau_4^{3/2}$. As shown in Fig.~\ref{ra_fig1}, oriented quivers need to be embedded in Calabi-Yau manifolds with two identical singularities which are exchanged by the orientifold involution. These singularities are obtained via D-term stabilisation which forces two del Pezzo divisors to shrink to zero size. The four-cycles in the geometric regime which are transversally invariant can be either del Pezzo divisors supporting non-perturbative effects or large cycles wrapped by a hidden D7 T-brane stack which is responsible for achieving a dS vacuum \cite{Cicoli:2015ylx}. 

On the other hand, Fig.~\ref{ra_fig2} and \ref{ra_fig3} show two different possible global embeddings of orientifolded quivers. In both cases the fractional D3 branes sit at an orientifolded singularity obtained by D-term fixing and the large cycle is wrapped by a hidden D7 T-brane stack. The only difference is in the behaviour under the involution of the two del Pezzo divisors in the geometric regime which are wrapped by ED3-instantons: in Fig.~\ref{ra_fig2} they are transversally invariant, and so they give rise to standard $O(1)$ instantons, while in Fig.~\ref{ra_fig3} they are exchanged under the involution, leading to a $U(1)$ instanton (for a review see \cite{Blumenhagen:2009qh}). Due to the technical difficulty to deal with $U(1)$ instantons, in this paper we shall focus only on the case depicted in Fig.~\ref{ra_fig2}. Due to the presence of two del Pezzo divisors in the geometric regime, such a model is also suitable to drive inflation, as we show in our explicit example: one of these blow-up modes can play the r\^ole of the inflaton while the other can keep the volume mode stable throughout the whole inflationary dynamics \cite{Conlon:2005jm}. Our model, therefore, represents an explicit Calabi-Yau compactification with both a chiral visible sector and a successful inflationary dynamics. 

This paper is organised as follows. In Sec.~\ref{sec:local-quiver} we describe the details of the local model and the corresponding orientifolded quiver while in Sec.~\ref{GEmb} we first list the consistency conditions for a successful global embedding and then we present a concrete Calabi-Yau example with an explicit choice of orientifold evolution, D-brane setup and gauge fluxes. Sec.~\ref{ModStab} provides a systematic analysis of all the effects which lead to full closed string moduli stabilisation in a Minkowski (or slightly dS) vacuum. In Sec.~\ref{secInfl} we then perform a complete multi-field analysis to show how our model can also successfully drive inflation. We finally present our conclusions in Sec.~\ref{Concl}.

\section{The local model}
\label{sec:local-quiver}

For concreteness, in this work we will focus on the field theories arising from D3-branes probing isolated orientifolds of the $\bC^3/\IZ_3$ orbifold. From the global embedding point of view, this kind of models are obtained by shrinking a dP$_0$ divisor to zero size. Higher order del Pezzo singularities can be considered in a similar way.

\subsection{Basics of $\bC^3/\IZ_3$ orientifolded quivers}

\begin{figure}
  \begin{subfigure}[b]{0.3\textwidth}
    \centering
    \includegraphics[width=\textwidth]{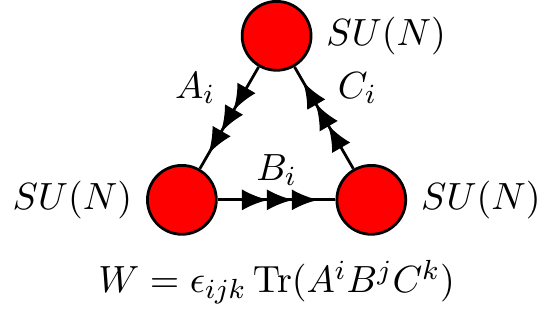}
    \caption{No projection.}
    \label{sfig:Z3-quiver}
  \end{subfigure}
  \hfill
  \begin{subfigure}[b]{0.3\textwidth}
    \centering
    \includegraphics[width=\textwidth]{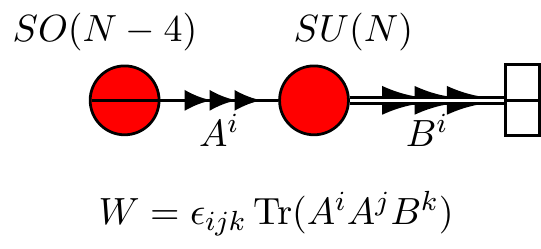}
    \caption{$SO$ projection.}
    \label{sfig:Z3-SO}
  \end{subfigure}
  \hfill
  \begin{subfigure}[b]{0.3\textwidth}
    \centering
    \includegraphics[width=\textwidth]{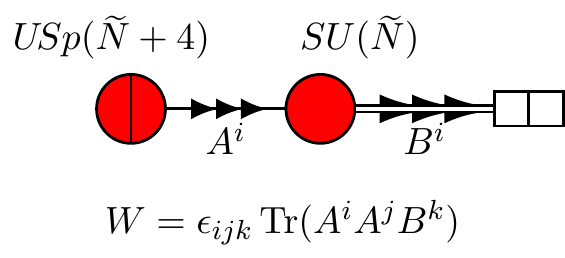}
    \caption{$\Sp$ projection.}
    \label{sfig:Z3-USp}
  \end{subfigure}
  \caption{\subref{sfig:Z3-quiver} Quiver for $N$ mobile D3 branes probing the $\bC^3/\IZ_3$ singularity, in the absence of orientifold projection. \subref{sfig:Z3-SO} and \subref{sfig:Z3-USp} The two possibilities for the theory after orientifolding via the projection described in the text.}
\end{figure}

In the worldsheet CFT describing type IIB on flat space, we gauge the group with generators $\{\Omega(-1)^{F_L}\cI,\cR\}$, with $\Omega$ worldsheet-parity, $F_L$ the left-moving fermion number, and the following geometric actions on the $\bC^3$ coordinates:
\begin{align}
  \cR& \colon (x,y,z) \to (\omega x, \omega y, \omega z)\, \\
  \cI& \colon (x,y,z) \to (-x, -y, -z)\, ,
\end{align}
where $\omega=\exp(2\pi i/3)$. Consider first the gauging of the $\cR$ generator, in the presence of D3-branes at $x=y=z=0$. The resulting quiver can be obtained by applying the techniques in \cite{Douglas:1996sw}, with the result being the quiver described in Fig.~\ref{sfig:Z3-quiver}. If we now gauge the $\Omega(-1)^{F_L}\cI$ generator we end up with the quiver theories depicted in Fig.~\ref{sfig:Z3-SO} and \ref{sfig:Z3-USp} \cite{Angelantonj:1996uy,Lykken:1997gy,Lykken:1997ub,Kakushadze:1998tr}. Which possibility we end up with depends on the choice of representation of the orientifold involution on the Chan-Paton factors.

More concretely, these theories are described by the following matter content, where we also indicate the transformation under the non-anomalous global symmetries. For the $\Sp$ projection we have:
\begin{align}
  \begin{array}{c|cc|ccc}
    & \Sp(N+4) & SU(N) & SU(3) & U(1)_R & \IZ_3 \\
    \hline
    A^i & \ov\fund & \fund & \fund & \frac{2}{3} - \frac{2}{N} & 1\\
    B^i & 1 & \ov\symm & \fund & \frac{2}{3} + \frac{4}{N} & -2
  \end{array}
\end{align}
while for the $SO$ projection we have:
\begin{align}
  \label{eq:Z3-SO-charges}
  \begin{array}{c|cc|ccc}
    & SO(N-4) & SU(N) & SU(3) & U(1)_R & \IZ_3 \\
    \hline
    A^i & \ov\fund & \fund & \fund & \frac{2}{3} + \frac{2}{N}& 1\\
    B^i & 1 & \ov\asymm & \fund & \frac{2}{3} - \frac{4}{N} & -2
  \end{array}
\end{align}
In both cases we have a non-vanishing superpotential of the form $W=\epsilon_{ijk}\Tr(A^iA^jB^k)$. The resulting brane system sources no D7 or D5 charge, while sourcing $(2N-3)/2$ units of mobile D3 charge (in the double cover) in the $SO$ case, and $(2N+3)/2$ units of mobile D3 charge in the $\Sp$ case \cite{Garcia-Etxebarria:2016bpb}.

Our goal in this paper is to study these theories as examples for semi-realistic string constructions, so choosing the $SO$ projection with $N=5$ would seem optimal: we obtain a $SU(5)$ theory with three generations of $\bf 5$ and $\overline{\bf{10}}$, reasonably reminiscent of conventional $SU(5)$ GUT models, except for the absence of the corresponding Higgs boson. Unfortunately, as we will review momentarily, the resulting theory has a dynamically generated runaway superpotential, which makes it unsuitable for model building purposes.\footnote{Moreover, the spectrum does not include the Standard Model Higgs field.}

\subsection{The $SU(5)$ model}

The non-perturbative dynamics of the $SU(5)$ case has been studied in \cite{Lykken:1998ec} (see also \cite{GarciaEtxebarria:2012qx} for a recent detailed study), which we now briefly review for completeness. (The contents of this section are not essential for the rest of the paper, and so can be safely skipped.)

Taking $N=5$ in~(\ref{eq:Z3-SO-charges}), and forgetting about the discrete global symmetry for simplicity, we have the field content: 
\be
  \begin{array}{c|c|cc}
     & \SU(5)  & \SU(3) & \U(1)_R \\ \hline
    A^i & \ov{\fund} & \fund & 16/15 \\
    B^i & \asymm & \fund & -2/15
  \end{array}
\ee
and superpotential:
\be
  \label{eq:classical-superpotential}
  W = \frac{1}{2} \lambda\, \epsilon_{i j k} A^i_m A^j_n B^{m n;\,
    k}\, .
\ee
It is convenient, as done in \cite{Lykken:1998ec}, to start by studying the theory in the absence of superpotential, i.e.\  sending $\lambda\to 0$. In this case there is an enhancement of the global symmetries, and we have:
\begin{equation}
  \begin{array}{c|c|cccc}
     & \SU(5)  & \SU(3) & \SU(3) & \U(1) & \U(1)_R \\ \hline
    A^i & \ov{\fund} & \fund & \rd1 & -3 & 16/15 \\
    B^i & \asymm & \rd1 & \fund & 1 & -2/15 \\ \hline \hline
    T_i^m = A^2 B & & \ov{\fund} & \fund & -5 & 2\\
    U^{i; m}_{\;\;n} = A B^3 & & \fund & \adj
     & 0 & 2/3\\
    V^{m n} = B^5     & & \rd1 & \symm & 5 & -2/3
  \end{array}
\end{equation}
We have also written the basic confined fields describing the IR of this s-confining theory \cite{Csaki:1996zb}. The $U(1)_R$ symmetry generator is chosen to agree with the one present in the theory with $\lambda\neq 0$. We have introduced:
\begin{eqnarray}
    T^m_i & \equiv & \frac{1}{2} \epsilon_{i j k} A^j_a A^k_b B^{a b ; m}\,,\qquad
    U^{i ; m}_{\;  \;  \; n} \equiv \frac{1}{12} \epsilon_{n p q}
    \epsilon_{b c d e f} A^i_a B^{a b ; p} B^{c d ; q} B^{e f ; m}\,, \\
    V^{m n} & \equiv & \frac{1}{160} \epsilon_{p q r} \epsilon_{a_1 b_1
      c_1 d_1 e_1} \epsilon_{a_2 b_2 c_2 d_2 e_2} B^{a_1 a_2 ; p} B^{b_1 c_1 ;
      q} B^{b_2 c_2 ; r} B^{d_1 e_1 ; m} B^{d_2 e_2 ; n}\,.
\end{eqnarray}
The confined description has a superpotential \cite{Csaki:1996zb}:
\be
  W = \frac{1}{\Lambda^9} \left(\epsilon_{m n p}\, T_i^m
    U^{i;n}_{\;\;q} V^{p q} - \frac{1}{3} \epsilon_{i j k}
    \,U^{i;m}_{\;\;p} U^{j;n}_{\;\;m} U^{k;p}_{\;\;n}\right)
\ee
whose equations of motion give the quantum moduli space, which in this case agrees with the classical moduli space.

We now reintroduce the superpotential coupling~(\ref{eq:classical-superpotential}), which in the confined variables becomes simply:
\be
W_{\rm tree} = \lambda T^i_i\, .
\ee
The full superpotential in the confined variables thus reads:
\be
W = \frac{1}{\Lambda^9} \left(\epsilon_{m n p}\, T_i^m
    U^{i;n}_{\;\;q} V^{p q} - \frac{1}{3} \epsilon_{i j k}
    \,U^{i;m}_{\;\;p} U^{j;n}_{\;\;m} U^{k;p}_{\;\;n}\right) + \lambda
  T^i_i\, .
\ee

It was shown in \cite{Lykken:1998ec} that, due to the linear term
proportional to $\lambda$, there is no solution to the F-term equations arising from this superpotential. As a simple illustrative example, it is clear that along the $\det(V)\neq 0$ directions one has a mass for the $T$ and $U$ fields, and upon integrating them out one obtains an effective superpotential given by:
\be
W_{\rm eff} = \frac{\lambda^3 \Lambda^{18}}{\det(V)}\, .
\label{Wrunaway}
\ee
This is perhaps best understood in terms of the S-dual configuration \cite{GarciaEtxebarria:2012qx}, where this effective superpotential arises from the ADS mechanism \cite{Affleck:1983mk}. The runaway directions caused by the superpotential (\ref{Wrunaway}) could be avoided by the presence of soft mass contributions for the matter fields generated by supersymmetry breaking background fluxes. However this open string stabilisation mechanism would lead to a complete breaking of the visible sector gauge group. Therefore we shall not consider this option but focus on visible sector configurations where no non-perturbative superpotential gets generated.

\subsection{The $SU(7)\times SO(3)$ model and beyond}
\label{Sec:SU7}

The theories that we are constructing have a conventional large $N$ dual description in terms of a freely acting orientifold of AdS$_5\times (S^5/\IZ_3)$, so for $N\geq N_\star$ we expect to have no runaway superpotential, as long as (the a priori unknown) $N_\star$ is large enough, since in these cases an interacting SCFT is expected to exist.

In general it is rather difficult to understand precisely the IR dynamics of the class of theories under consideration, so the determination of $N_\star$ is quite non-trivial, but in this case  a shortcut exists: the arguments of \cite{GarciaEtxebarria:2012qx} imply (assuming that the duality proposed in that paper is correct) that the $N=7$ theory (i.e. the $SU(7)\times SO(3)$ theory) already has a supersymmetric minimum. This follows since the strong coupling dual of this theory, which has $\Sp(8)\times SU(4)$ gauge group, can be shown to become free in the IR. The same should then be true of the
$SU(7)\times SO(3)$ theory. Thus, in order to prevent any runway in the open string sector, one simple way out of our predicament is to take $N\geq N_\star =7$ (that is, adding extra D3-branes), while keeping the rest of the setup untouched. This is what we will assume for the rest of the paper.

\section{Global embedding}
\label{GEmb}

\subsection{General consistency conditions}

As mentioned in the introduction, there is a technical challenge to embed local quiver models in fully consistent string compactifications. Given a concrete local model, there may be several ways to embed it into the myriad of Calabi-Yau compactifications known to date that have the corresponding singularity. The consistency conditions are:
\begin{description}
\item {\it Global embedding of the orientifolded singularity}: The local model consists of a point-like singularity and an orientifold involution that makes the singular point fixed under the involution itself. The embedding of the local model requires to find a Calabi-Yau that can admit the desired singularity and allows for a globally defined involution which keeps the singular point fixed. This is a non-trivial task. Moreover, there are two possibilities that could
  arise: 1) the singular point is an isolated fixed point, or 2) it sits on top of a fixed codimension one locus (i.e.\ divisor). The case discussed in the previous section is of the first kind, and this is what we will focus on in this paper. The second possibility is also interesting but introduces the difficulty of having O7-planes extending into the bulk, which slightly complicates the construction of tadpole-free global models.

\item {\it D7-tadpole cancellation}: The introduction of an orientifold involution generates a set of O-planes. These O-planes have a non-zero D7-charge which needs to be cancelled on a compact space. Hence, we have to introduce some D7-branes in the background. The easiest choice to cancel the D7-tadpole is to put 4 D7-branes (plus their 4 images) on top of the O7-locus. This will generate a hidden sector  with $SO(8)$ gauge group. As we will see, in some situations a two-form flux must be switched on along the D7-brane worldvolume. 

\item {\it D3-tadpole cancellation}: Both the D-branes at the singularity, the hidden sector D7-branes (with or without fluxes) and the O-planes carry a net D3-charge. Summing all their contribution typically gives a negative number (if the gauge flux does not contribute with a positive large number). This needs to be cancelled by other contributions coming from different objects in the compactification, i.e.\ mobile D3-branes and 3-form fluxes. These last ones in fact carry a positive D3-charge. Having a large negative D3-charge coming from the D-brane setup is desirable in order to have a large D3-charge at our disposal for switching on a large number of tunable 3-form fluxes (necessary to fix the complex structure moduli and the axio-dilaton).

\item {\it Non-perturbative effects}: In order to stabilise the K\"ahler moduli, some non-perturbative effects should be present. One typically needs rigid cycles in the compact Calabi-Yau that can support $O(1)$ instantons (or D7-brane stacks undergoing gaugino condensation) generating a non-perturbative superpotential $W_{\rm np}\sim A\,e^{-aT}$ (where $T$ is the K\"ahler modulus whose real part measures the size of the rigid divisor). The presence of D7-branes in the compactification can spoil the generation of such a non-perturbative superpotential and one needs to constrain the D7-brane data to avoid this clash \cite{Blumenhagen:2007sm}. The best situation is when non-perturbative effects are supported on rigid del Pezzo divisors which do not intersect with the visible sector D-brane stack \cite{Cicoli:2011qg}.
\end{description}

In the following, we will present an explicit, consistent, global setup where all these issues are taken into account.

\subsection{Explicit example: global orientifolded dP$_0$}

Let us consider the Calabi-Yau three-fold $X$ defined by the following toric-data \cite{Diaconescu:2007ah}:\footnote{Each line corresponds to a $\mathbb{C}^\ast$ action. The last column gives the degrees of the anti-canonical class of the toric ambient space.} 
\begin{equation}
\begin{array}{|c|c|c|c|c|c|c|c||c|}
\hline W_1 & W_2 & W_3 & W_4 & W_5 & Z & X & Y & D_\textmd{X} \tabularnewline \hline \hline
    0  &  0  &  0  &  0  &  0  &  1  &  2  &  3  & 6\tabularnewline\hline
    1  &  1  &  1  &  0  &  0  &  0  &  6  &  9  & 18\tabularnewline\hline
    0  &  1  &  0  &  1  &  0  &  0  &  4  &  6  & 12\tabularnewline\hline
    0  &  0  &  1  &  0  &  1  &  0  &  4  &  6  & 12\tabularnewline\hline
\end{array}\label{eq:model3dP8:weightm}\,,
\end{equation}
and Stanley-Reisner ideal:
\be
\label{eq:model3dP8:sr-ideal}
\begin{split}
{\rm SR}=\{W_1\, W_2\,W_3,\,& W_2\, W_4,\, W_3 \, W_5,\, W_4\, W_5, \\
& W_1\,W_2\,X\,Y, \, W_1\,W_3\,X\,Y, \,  W_4\,Z, \,W_5\,Z, \, X\, Y\, Z\}\,.
\end{split}
\ee
The Calabi-Yau $X$ is a hypersurface in the above ambient space given by the vanishing of a polynomial whose degrees can be read from the last column of (\ref{eq:model3dP8:weightm}). Its Hodge numbers are $h^{1,1}(X)=4$ and $h^{1,2}(X)=214$ (the Euler characteristic is then $\chi=2(h^{1,1}-h^{1,2})=-420$).

A basis of $H^{1,1}(X)$ is given by:\footnote{This is not an {\it integral basis} (i.e. a basis such that any integral divisor is a linear combination of the basis elements with integral coefficients): for example $D_{W_1}=\frac13(\cD_1-3\cD_2-3\cD_3-\cD_4)$.}
\be
\label{dP8model-basis}
\cD_1 = 3D_{W_3} + 3D_{W_4} + D_{Z} \qquad \cD_2 = D_{W_4} \qquad \cD_3 = D_{W_5} \qquad \cD_4 =  D_{Z} \:.
\ee
The intersection form in this basis is diagonal:
\be
\label{IntersExample2}
I_3 = 9\cD_1^3 + \cD_2^3 + \cD_3^3 + 9\cD_4^3 \:.
\ee
This Calabi-Yau threefold has one dP$_0$ at $Z=0$ and two dP$_8$'s at $W_4=0$ and $W_5=0$. The second Chern class of the Calabi-Yau threefold is:\footnote{In this paper, we use the same symbols for the divisors of $X$ and their Poincar\'e dual two-forms.}
\be
c_2(X) = \frac{1}{3} \left( 34\, \cD_1^2 +30\, \cD_2^2 +30\, \cD_3^2 -2\, \cD_4^2 \right) \:,
\ee
where the $1/3$ factor appears because we are not using an {\it integral basis} (but $c_2(X)$ is an integral four-form). Its simple form is due to the intersections (\ref{IntersExample2}) in the chosen basis.

Expanding the K\"ahler form in the basis (\ref{dP8model-basis}), $J=\sum_i t_i\cD_i$, one has the following volumes of the three del~Pezzo divisors:
\be
{\rm Vol}(D_Z) \equiv \tau_4 = \tfrac92 t_4^2 \:, \qquad
{\rm Vol}(D_{W_4}) \equiv \tau_2 = \tfrac12 t_2^2 \:,\qquad
{\rm Vol}(D_{W_5}) \equiv \tau_3 = \tfrac12 t_3^2 \:,
\ee
and the volume of the Calabi-Yau three-fold is:
\be
{\rm Vol}(X) \equiv \vo = \frac16 (9 t_1^3 + t_2^3 + t_3^3 + 9 t_4^3) \:.
\ee
The K\"ahler cone of the ambient space is:
\be
t_2<0 \qquad t_3<0 \qquad t_1+t_2+t_4>0 \qquad t_1+t_3+t_4>0 \qquad t_4<0 \:.
\label{KC}
\ee
This space is a priori only a subspace of the K\"ahler cone of the Calabi-Yau. However, the point we want to consider, i.e.\ a Calabi-Yau with two finite size dP$_8$ divisors and one dP$_0$ singularity, is included in this subspace (at the boundary).\footnote{When the dP$_0$ shrinks, i.e. for $t_4\rightarrow 0$, the K\"ahler cone becomes: \be
t_2<0 \qquad t_3<0 \qquad t_1+t_2>0 \qquad t_1+t_3>0  \,. \nn
\ee}

Given the K\"ahler cone conditions in (\ref{KC}), the overall volume in terms of the four-cycle moduli looks like:
\be
\vo = \frac{\sqrt{2}}{9}\left(\tau_1^{3/2} - 3\tau_2^{3/2} - 3\tau_3^{3/2} - \tau_4^{3/2}\right)\,.
\label{CYvo}
\ee

\subsubsection{Orientifold involution}

The Calabi-Yau at hand has only one involution coming from the toric variety which has $Z=0$ as a fixed point set, i.e:
\be
\label{Zinvol}
\sigma\,:\qquad Z\rightarrow-Z \:.
\ee
The hypersurface equation that respects this involution takes the form:\footnote{We allow only monomials with even powers of $Z$.}
\be
\label{eq:hse-second-ex-restricted}
\textmd{eq}_{\tilde X}=Y^2+X^3+\sum_{n=1}^3 A_{0,6n,4n,4n}(W_i)\, X^{3-n}\, Z^{2n}=0\,,
\ee
where $A_{0,6n,4n,4n}(W_i)$ are polynomials in $W_i$ with the indicated degree. Although it might look as if (\ref{eq:hse-second-ex-restricted}) describes a non-generic Weierstra\ss{} fibration, we see from the SR-ideal (\ref{eq:model3dP8:sr-ideal}) that the fibration structure is not respected by the triangulation of the polytope. Otherwise, the divisor $D_Z$ would be a dP$_2$. 

The fixed point set of the involution (\ref{Zinvol}) is given by the codimension-1 loci $\{Z=0\}$ and $\{Y=0\}$ and two fixed points $\{W_1=W_3=W_4=0\}$ and $\{W_1=W_2=W_5=0\}$. The last two loci are O3-planes (each contributing with $-1/2$ to the D3-charge).

\subsection{D-brane setup}

After shrinking the dP$_0$ surface at $Z=0$ (which, as we will show in Sec.~\ref{secD}, can be induced by D-term stabilisation), we obtain a singular point that is left fixed by the orientifold involution. Placing D3-branes on top of it generates the quiver gauge theory described in Sec.~\ref{Sec:SU7}.

We will cancel the D7-tadpole generated by the O7-plane at $Y=0$ by putting four D7-branes (plus their four images) on top of $Y=0$. This will give the hidden sector responsible for achieving a de Sitter vacuum.

Since there are two rigid dP$_8$ divisors that are invariant under the orientifold involution, they will be wrapped by $O(1)$ ED3-instantons. The wrapping number can be $1$ if the gauge invariant flux $\cF=F-\iota^\ast B$ can be set to zero. In order to have this, we will choose the $B$-field to be:\footnote{This choice is not necessary if we allow ED3-instantons wrapping several times the invariant divisor \cite{Berglund:2012gr}. On the other hand, a wrapping number bigger than $1$ would make it difficult to fix the volume to a value that is not too large \cite{Cicoli:2011qg}.}
\be
\label{BfieldEx2}
B = \frac{\cD_2}{2} + \frac{\cD_3}{2} \:.
\ee
This allow us to set $\cF_{E3_2}=\cF_{E3_3}=0$ by the proper choice of the half-integral gauge fluxes $F_{E3_2}$ and $F_{E3_3}$ on the two ED3-instantons.

We would also like to have zero chiral intersections between the ED3-instantons, wrapping $\cD_2$ and $\cD_3$, and the D7-branes wrapping the divisors $D_Y=3\cD_1-3\cD_2-3\cD_3$ (with Euler characteristic $\chi(D_Y)=c_2(X)D_Y+D_Y^3=435$). This can be done by properly choosing the flux on the D7-branes. The chiral intersections are given by:
\be
I_{E3_i \,D7} = \int_{\cD_i\cap D_Y} \cF_{D7} - \cF_{E3_i} = \int_{\cD_i\cap D_Y} \cF_{D7} \qquad \mbox{with }\alpha=2,3 \,, \nonumber
\ee
where in the last equality we have used $\cF_{E3_i}=0$. The gauge flux on the D7-brane must be properly quantised to cancel the Freed-Witten anomaly \cite{Freed:1999vc}. In the present case:
\be
F_{D7} + \frac{\iota^\ast\cD_1 }{2}+ \frac{\iota^\ast\cD_2 }{2}+ \frac{\iota^\ast\cD_3}{2}\in H^2(D_Y,\mathbb{Z}) \:,
\ee
where $\iota^\ast D$ is the pull-back of the CY two form $D$ on the D7-brane worldvolume.
A flux satisfying this condition is of the form:
\be
\label{FD7Ex2}
F_{D7} =  \iota^\ast F_{D7}^{\rm int}+\frac{\iota^\ast\cD_1 }{2}+ \frac{\iota^\ast\cD_2 }{2}+ \frac{\iota^\ast\cD_3}{2}\:,
\ee
where $F_{D7}^{\rm int}$ is an integral two-form of $X$, i.e.\ it is given in terms of the integral basis $\{D_{W_1},\cD_2,\cD_3,\cD_4\}$ by:
\be
F_{D7}^{\rm int} = n_{W_1}D_{W_1} + n_2 \cD_2 + n_3 \cD_3 + n_4 \cD_4 \qquad \mbox{with} \qquad n_i\in \mathbb{Z}\:.
\ee
Recall that $D_{W_1}=\frac13 (\cD_1 - \cD_4) - \cD_2 - \cD_3$. In terms of the integers $n_i$, the constraints $I_{E3_i\,D7}=0$ become:
\be
 n_{W_1} = n_2 = n_3  = n \qquad \mbox{for arbitrary integer }n\:.
\ee
The integer $n_4$ does not enter in the constraints, as $\iota^\ast \cD_4=0$ on the surface $Y=0$. The gauge invariant flux on the D7-brane depends on one integer number $n$ as:
\be
\label{FluxChoiceIImodel}
\cF_{D7} = \left( \frac{n}{3} + \frac12 \right) \iota^\ast \cD_1 \:.
\ee
The D3-charge generated by this flux is:
\be
Q_{D3}^{\cF_{D7}} = - 8 \times \frac12 \int_{D_Y} \cF_{D7}\wedge \cF_{D7} = - 3 (2n+3)^2 \:. 
\ee
The fact that this is negative means that the flux on the D7-brane will never be supersymmetric in the absence of a non-vanishing vacuum expectation value (VEV) for an open string scalar field $\phi$.\footnote{In fact, a supersymmetric anti-self-dual flux would give a positive contribution to $Q_{D3}^{\cF_{D7}}$.}

Before switching on the flux $\cF_{D7}$, the fields living on the worldvolume of the D7-brane stack are an 8D gauge connection and a scalar field $\Phi$, both in the adjoint representation of $SO(8)$. In this section we study what is the 4D effective field theory around a zero VEV for the D7-brane worldvolume scalar field when we switch on the flux (\ref{FluxChoiceIImodel}). This non-zero flux breaks the gauge group from $SO(8)$ to $U(4)$ and it generates a zero mode spectrum in the antisymmetric representation of $U(4)$ whose net chirality is given by $ I_{U(4)}^A = \frac12 I_{D7-D7'} + I_{D7-O7}$ where:
\be
I_{D7-D7'} =  \int_{D7\cap D7'}\cF_{D7}-\cF_{D7'}\qquad\mbox{and} \qquad  I_{D7-O7} =  \int_{D7\cap O7}\cF_{D7}  \:. 
\ee
In our case $[D7]=[D7']=[O7]=D_Y$ and $\cF_{D7'}=-\cF_{D7}$. Hence the number of chiral zero modes is:
\be
\label{netchiralityphi}
 I_{U(4)}^A = 2\int_X D_Y \wedge D_Y \wedge \cF_{D7} = 27(2n+3) \:.
\ee
A non-zero gauge flux on the D7-stack generates also a moduli-dependent Fayet-Iliopoulos (FI) term associated with the diagonal $U(1)$ of $U(4)$:
\be
\xi_{D7} = \frac{1}{4\pi\cV} \int_{D_Y} J\wedge \cF_{D7} = \frac{I^A_{U(4)}}{24\pi} \, \frac{t_1}{\vo}\simeq \frac{I^A_{U(4)}}{24\pi} 
\left(\frac23\right)^{1/3}\frac{1}{\vo^{2/3}}\,,
\label{FI}
\ee
where we have expanded the K\"ahler form as $J=\sum_i t_i \cD_i$ and we have approximated the overall volume as $\vo\simeq \frac{\sqrt{2}}{9} \tau_1^{3/2}$.

\subsection{T-brane background}

Above we have studied the effective theory on the D7-brane stack around the background where the D7-branes are on top of the O7-locus $D_Y$ and have zero VEV for the adjoint complex scalar $\Phi$ living on the D7-brane worldvolume and in the adjoint representation of $SO(8)$. We have seen that switching on a gauge flux on the D7-branes, the gauge group is broken to $U(4)$. However, the supersymmetric equation of motion for the 8D theory when the flux (\ref{FluxChoiceIImodel}) is turned on looks like: 
\be
J\wedge \cF_{D7} + \left[\Phi,\Phi^\dagger\right] dvol_4 = 0 \,.
\label{SUSY8D}
\ee
Given that $\cF_{D7}$ is never a primitive two-form for a non singular $J$, (\ref{SUSY8D}) forces $\Phi$ to develop a proper non-zero VEV. Hence $\Phi=0$ is not the true vacuum which is instead characterised by a non-vanishing VEV of both the gauge connection and the adjoint scalar field $\Phi$. This solution consists of a so-called T-brane background with a given D3-charge which gives rise to a non-Abelian gauge group without any $U(1)$ factor. Let us notice that the 8D supersymmetric condition (\ref{SUSY8D}) corresponds to the vanishing of the D-term potential from the 4D point of view. As we shall see in Sec.~\ref{secD}, the fact that the FI-term in (\ref{FI}) can never be zero for a finite Calabi-Yau volume forces some zero modes of $\Phi$ to get a non-zero VEV, leading the vacuum solution away from $\Phi=0$. 

Let us now describe this background solution more in detail. After the breaking $SO(8)\rightarrow U(4)$, the adjoint representation of $SO(8)$ is broken as:
\be
{\bf 28} \rightarrow {\bf 16}_0 \oplus {\bf 6}_{+2} \oplus {\bf 6}_{-2} \,,
\label{phicharge}
\ee
where ${\bf R}_q$ is in the representation ${\bf R}$ for $SU(4)$ and has charge $q$ with respect to the diagonal $U(1)$. Accordingly the scalar field $\Phi$ can be written as:\footnote{We exchange the first four rows with the second four rows with respect to the usual matrix notation for the adjoint of $SO(8)$.}
\be
\label{Ex2PhiSO8}
 \delta\Phi = \left(\begin{array}{cc}
     \phi_{{\bf 16}_0} & \phi_{{\bf 6}_{+2}} \\   \phi_{{\bf 6}_{-2}} &  -\phi_{{\bf 16}_0}^T \\
 \end{array}\right) \:.
\ee
In this basis, the first four lines (and columns) refer to the four D7-branes, while lines (and columns) from the fifth to the eighth refer to their images. Hence the upper right block corresponds to strings going from the four D7-branes to their images, while the lower left block corresponds to strings with opposite orientation (in fact, they have opposite charges with respect to the diagonal $U(1)$). Giving a VEV to both $\phi_{{\bf 6}_{+2}}$ and $\phi_{{\bf 6}_{+2}}$ recombines some of the four D7-branes with some of the image D7-branes. On the other hand, $\phi_{{\bf 16}_0}$, that is in the adjoint of $U(4)$,
describes deformations and the recombinations of the $U(4)$ stacks (that will be accompanied by the same process in the image stack).

These deformations of the theory can be studied globally, by reading the value of $\Phi$ from the tachyon matrix describing the D7-brane configuration. Let us describe this in a simple example. Let us consider a stack of D7-branes on top of a divisor $D_{z}=\{z=0\}$. Their configuration is described by the vanishing of the polynomial $z^2=0$. This setup, including also the possible gauge flux is described by the tachyon matrix \cite{Sen:1998sm,Witten:1998cd,Katz:2002gh,Aspinwall:2004jr,Collinucci:2008pf}:
\be
T = \left(\begin{array}{cc}
z&0\\0&z\\
\end{array} \right)
\ee
that is a map between two vector bundles \cite{Moore:2003vf}:
\be
 T :  \qquad \begin{array}{c}
\sG (-\frac{D_z}{2}+F_1)\\ \oplus \\ \sG (-\frac{D_z}{2}+F_2) \\
\end{array}  \qquad \rightarrow \qquad \begin{array}{c}
\sG (\frac{D_z}{2}+F_1)\\ \oplus \\  \sG (\frac{D_z}{2}+F_2) \\
\end{array} 
\ee
The line bundles on the left are related to anti-D9-branes while the ones on the right are related to D9-branes. The two-forms $F_1$ and $F_2$ are arbitrary.
The tachyon condensation will produce the annihilation of the D9 and the anti-D9-branes wherever $T$ has full rank and  is therefore a bijective map. On the other hand, something remains on the locus where $T$ has lower rank. In this case, we see that $\det T=z^2=0$. Hence on $z^2=0$ the rank of T decreases and we are left with two D7-branes. The total D-brane charge is conserved in this process and hence can be computed before the tachyon condensation: 
\be
\label{eq:charge-vector}
\Gamma= \left(  \mbox{ch}(D9) - \mbox{ch}(\overline{D9}) \right) \left( 1+ \frac{c_2(X)}{24} \right)\,.
\ee 
In our case $\mbox{ch}(D9) = e^{\frac{D_z}{2}+F_1}+e^{\frac{D_z}{2}+F_2}$, while $\mbox{ch}(\overline{D9}) = e^{-\frac{D_z}{2}+F_1}+e^{-\frac{D_z}{2}+F_2}$. Plugging these expressions into (\ref{eq:charge-vector}), one obtains the charge vectors of two D7-branes wrapping the divisor $D_z$, one with flux $F_1$ and the other with flux $F_2$.

If we now deform $T$ by adding for example:
\be
\Phi = \left(\begin{array}{cc}
z_1&0\\0&z_2\\
\end{array} \right) \:,
\ee
then det$(T+\Phi)=(z+z_1)(z+z_2)$, i.e.\ the two branes split into two (almost everywhere) separated D7-branes. If we instead deform $T$ by adding:
\be
\label{PhiRecomb}
\Phi = \left(\begin{array}{cc}
0&x_1\\x_2&0\\
\end{array} \right) \:,
\ee
then det$(T+\Phi)=z^2- x_1x_2$, i.e.\ the two D7-brane have recombined into one D7-brane. On the other hand if we set $x_2\equiv 0$ in (\ref{PhiRecomb}) while keeping $x_1\neq 0$, the equation defining the D7-brane configuration is $z^2=0$, i.e.\ the same as of a stack of two D7-branes. However, the gauge group is broken from $U(2)$ to $U(1)$. This is a T-brane \cite{Gomez:2000zm,Donagi:2003hh,Donagi:2011jy,Cecotti:2010bp} (the name is based on  the triangular form of (\ref{PhiRecomb}) when $x_2\equiv 0$) background: the two D7-branes form a bound state, whose gauge group is $U(1)$ and whose tachyon matrix is:
\be
T = \left(\begin{array}{cc}
z&x_1\\0&z\\
\end{array} \right) \:.
\ee
This tachyon matrix (with its domain and codomain) is the only information we need to calculate the D-brane charges of the T-brane.

In an orientifolded theory, where the orientifold involution acts as $\xi\mapsto -\xi$ for some coordinate $\xi$, the full tachyon (describing the invariant D7-brane configuration that cancels the O7-plane tadpole) must satisfy the condition \cite{Collinucci:2008pf}:
\be
 T = \xi S + A\,,
\ee
where $S$ is a symmetric matrix and $A$ an antisymmetric one in the standard basis of \cite{Collinucci:2008pf}. There exists a change of basis, such that  the matrices  $S$ and $A$ take the following form:
\be
S = \left(\begin{array}{cc}
M_S&S_1\\S_2&M_S^T\\
\end{array} \right) \qquad\mbox{ and }\qquad 
A = \left(\begin{array}{cc}
M_A&A_1\\A_2&-M_A^T\\
\end{array} \right) \:,
\ee
where $M_{S,A}$ are generic $N\times N$ matrices, $S_{1,2}$ are symmetric $N\times N$ matrices and $A_{1,2}$ are antisymmetric $N\times N$ matrices.
In this basis, the first $N$ lines (and columns) refer to a set of $N$ branes, while the last $N$ lines (and columns) refer to their $N$ images.

Let us apply this formalism to our setup, where we have an orientifold plane at $Y=0$ and four D7-branes (plus their four images) on the same locus. The tachyon of this configuration is very simple and it is given by (in our case $Y=\xi$):
\be
\label{TSO8}
T = \left(\begin{array}{cc}
Y {\bf 1}_4& 0 \\ 0 & Y {\bf 1}_4 \\
\end{array} \right) \:.
\ee
We need to specify also the domain and codomain of this map. As we have seen in the toy example above, this will determine the flux on the T-brane background. In the chosen setup, all the four D7-branes have the same flux. This is realised by the following map \cite{Collinucci:2008pf,Collinucci:2010gz}:
\be
\label{TSO8domcodom}
T :  \qquad \begin{array}{c}
\sG (-\frac{D_Y}{2}-F_{D7}+2B)^{\oplus 4}\\ \oplus \\ \sG (-\frac{D_Y}{2}+F_{D7})^{\oplus 4} \\
\end{array}  \qquad \rightarrow \qquad \begin{array}{c}
\sG (\frac{D_Y}{2}-F_{D7}+2B)^{\oplus 4}\\ \oplus \\  \sG (\frac{D_Y}{2}+F_{D7})^{\oplus 4} \\
\end{array} 
\ee
where $F_{D7}$ and the $B$-field are defined in (\ref{FD7Ex2}) and (\ref{BfieldEx2}).\footnote{The orientifold symmetry imposes constraints also on domain and codomain, that include also the $B$-field.}

We can deform the background (\ref{TSO8}) by adding a matrix of the form (\ref{Ex2PhiSO8}). As we said we are interested in switching on only a off-diagonal block in (\ref{Ex2PhiSO8}). One can switch on both off-diagonal blocks only when the gauge invariant flux $\cF_{D7}=F_{D7}-B$ satisfies:
\be
-\frac{D_Y}{2} \leq \cF_{D7} \leq \frac{D_Y}{2} \:.
\ee
Otherwise, one of the off-diagonal block does not have the degrees necessary for a holomorphic section. In our setup this allows only the following values for $n$:
\be
-\frac92 \leq n \leq \frac32\,, \qquad \mbox{ i.e. } \qquad n=-4,-3,-2,-1,0,1\,.
\label{possibn}
\ee
The supersymmetric constraint (\ref{SUSY8D}) tells us which block we have to switch on. One obtains the same result by studying the stability condition for the D-branes (see \cite{Collinucci:2014qfa}). In any case, the deformation does not change the D-brane charge, that can be read off from (\ref{TSO8domcodom}):
\bea
\Gamma_{D7} &=& 4 e^{-B} \,\left(  e^{\frac{D_Y}{2}-F_{D7}+2B} + e^{\frac{D_Y}{2}+F_{D7}} - e^{-\frac{D_Y}{2}-F_{D7}+2B} - e^{-\frac{D_Y}{2}+F_{D7}}
\right) \left( 1+ \frac{c_2(X)}{24} \right) \nonumber \\
  &=&  24(\cD_1 - \cD_2 - \cD_3) +\frac13 \left[ -57 (\cD_2^3 + \cD_3^3) + 2\cD_1^3 (35 + 6 n + 2 n^2) \right]\, .
\eea
The D-brane charge of the O7-plane at $Y=0$ is:
\be
\Gamma_{O7} = -8 D_Y + D_Y\frac{D_Y^2+c_2(X)}{6} = - 24(\cD_1 - \cD_2 - \cD_3) + \frac16 \left[61 \cD_1^3 - 57 (\cD_2^3 + \cD_3^3) \right]\:.
\ee
Summing them together, we actually see that all charges cancel except the D3-charge:
\be
\Gamma = \Gamma_{D7}+\Gamma_{O7} = \frac16 \left[-171 (\cD_2^3 + \cD_3^3) + \cD_1^3 (201 + 24 n + 8 n^2)\right]\:.
\ee
Now we can compute:
\be
Q_{D3} = - \int_X \Gamma|_{\rm 6-form} = -\frac32 (163 + 24 n + 8 n^2)\,
\ee
which for $n$ given by (\ref{possibn}) takes the three possible values $Q_{D3}=-\frac{585}{2}, -\frac{489}{2}, -\frac{441}{2}$. This charge is half integral. To compute the total D3-charge, we need to add the half-integral D3-charge of the fractional branes, that for $N=7$ is equal to $\frac{11}{2}$, and the charge of the two O3-planes (each one equal to $\frac12$), i.e.  $Q_{D3}^{\rm tot}= -286, -238, -214$. This large negative D3-charge allows for a large tunability of the fluxes (that have typically positive D3-charge).

In our situation we have the following T-brane solution:
\be
 \langle\Phi\rangle = \left(\begin{array}{cc}
   0 & \Phi_{{\bf 6}_{+2}}   \\ 0 & 0 \\
 \end{array}\right) \:.
\ee
The locus where the brane are sitting is still the same as for $\langle\Phi\rangle=0$, but the gauge group is reduced to $Sp(1)\times SO(4)$ without any $U(1)$ factor. As we shall see in Sec.~\ref{secF} a non-zero VEV for $\Phi$ produces couplings in the 4D EFT after compactification which are crucial to obtain dS vacua \cite{Cicoli:2015ylx}. 

\section{Moduli stabilisation}
\label{ModStab}

We will now describe how all closed string moduli can be stabilised in a dS vacuum. These consist of $h^{1,2}_-$ complex structure moduli $U_\alpha$, the axio-dilaton $S =g_s^{-1} + {\rm i} C_0$ and $h^{1,1}_+$ K\"ahler moduli defined as $T_i = \tau_i + {\rm i} \rho_i$ where the axions are given by $\rho_i=\int_{\cD_i} C_4$. In our example $h^{1,1}_-=0$ ($h^{1,1}_-$ counts the number of $(B_2,C_2)$-axions) and $h^{1,2}_+=0$ ($h^{1,2}_+$ counts the number of bulk $U(1)$'s). Hence we will have $h^{1,2}_-=h^{1,2}(X)=214$ and $h^{1,1}_+=h^{1,1}(X)=4$.

\subsection{Background fluxes and D-terms}
\label{secD}

We shall now show how to fix all these moduli, realising an explicit LARGE Volume Scenario (LVS) \cite{Balasubramanian:2005zx, Conlon:2005ki, Cicoli:2008va}. Given that the overall volume $\vo$ turns out to be exponentially large in string units, the contributions to the 4D scalar potential from different sources can be effectively organised in a $1/\vo\ll 1$ expansion. The leading terms emerge at $\mc{O}(\vo^2)$ and they arise from three-form background fluxes $G_3$ and D-terms. The tree-level K\"ahler potential $K$ and the superpotential $W$ read \cite{Gukov:1999ya} (setting $M_p=1$):
\be
K = - \ln\left(S+{\bar S}\right) -\ln\left(-i\int_X\Omega (U) \wedge{\bar\Omega}\right)-2\ln\vo \qquad\qquad W=\int_X G_3 \wedge \Omega\,.
\label{Ktree}
\ee
Due to the no-scale cancellation, the supergravity F-term scalar potential takes the simple form:
\be
V_F^{\rm flux}\simeq \frac{1}{\vo^2}\left(|D_S W|^2+\sum_{\alpha=1}^{h^{1,2}_-} |D_{U_\alpha} W|^2\right)\,.
\ee
This expression is positive semi-definite and admits a Minkowski minimum at $D_S W = D_{U_\alpha} W = 0$ where the axio-dilaton and all complex structure moduli are fixed supersymmetrically. Supersymmetry is instead broken along the K\"ahler moduli directions which are however still flat at this order of approximation. 

Other contributions of $\mc{O}(1/\vo^2)$ come from D-terms associated with the anomalous $U(1)$'s living respectively on the D7-stack wrapped around the O7-plane and the D3-brane at the dP$_0$ singularity. Thus the total D-term potential is given by:
\be
V_D = V_D^{\rm bulk} + V_D^{\rm quiver}\,.
\ee
The bulk D-term potential reads (we are following the same conventions as \cite{Cicoli:2015ylx}): 
\be
V_D^{\rm bulk} =\frac{1}{2{\rm Re}(f_{D7})} \left(\sum_i q_{\phi_i} \frac{|\phi_i|^2}{{\rm Re}(S)}  - \xi_{D7}\right)^2 \,,
\ee
where $\xi_{D7}$ is given in (\ref{FI}), $q_{\phi_i}$ are the $U(1)$ charges of the canonically unnormalised fields $\phi_i$ and the hidden sector gauge kinetic function is $f_{D7} = 3(T_1-T_2-T_3)/(2\pi)$. Considering, without loss of generality, just one canonically normalised charged matter field $\varphi$ and approximating ${\rm Re}(f_{D7})\simeq 3 \tau_1 /(2\pi) \simeq \frac{9}{2\pi} \left(\frac 32\right)^{1/3} \vo^{2/3}$, the bulk D-term potential becomes:
\be
V_D^{\rm bulk} = \frac{c_1}{\vo^{2/3}} \left(q_\varphi |\varphi|^2 - \frac{c_2}{\vo^{2/3}} \right)^2\,.
\label{VDbulk}
\ee
where:
\be
c_1 = \frac{\pi}{9}\left(\frac 23\right)^{1/3}\qquad\text{and}\qquad c_2 = \frac{I^A_{U(4)}}{24\pi}\left(\frac23\right)^{1/3}\,.
\ee 
Clearly (\ref{VDbulk}) scales as $\mc{O}(1/\vo^2)$ (as can be seen by setting $\varphi=0$). On the other hand, the quiver D-term potential takes the form:
\be
V_D^{\rm quiver} =\frac{1}{2{\rm Re}(f_{D3})} \left(q_C |C|^2  - \xi_{D3}\right)^2 \,,
\label{VDquiver}
\ee
where again, without loss of generality, we focused on just one canonically normalised visible sector matter field $C$; the gauge kinetic function at the quiver singularity is $f_{D3} = S/(2\pi)$ and the visible sector FI-term scales as \cite{Cicoli:2012vw, Cicoli:2013mpa, Cicoli:2013zha, Cicoli:2013cha}:
\be
\xi_{D3} \simeq \frac{\tau_4}{\vo}\,.
\ee
Again (\ref{VDquiver}) clearly scales as $\mc{O}(1/\vo^2)$. Given that the two D-term potentials are positive semi-definite, both of them are minimised supersymmetrically at $V_D^{\rm bulk}=V_D^{\rm quiver}=0$. These conditions fix the combinations of moduli corresponding to the combinations of closed and open string axions which get eaten up by the two anomalous $U(1)$'s. Given that these combinations are mostly given by an open string axion for branes wrapping cycles in the geometric regime, while they are mostly given by closed string axions for branes at singularities \cite{Allahverdi:2014ppa}, the moduli fixed by the D-terms are:
\be
|\varphi|^2 = \frac{c_2}{q_\varphi \vo^{2/3}} \qquad\text{and}\qquad \tau_4 \simeq q_C |C|^2 \vo\,.
\label{Dfix}
\ee
Given that the string scale $M_s$ is written in terms of the Planck scale as $M_s\sim M_p/\sqrt{\vo}$, the leading order potential generated by background fluxes and D-terms scales as $M_s^4$. It is therefore crucial that this potential vanishes at the minimum since otherwise we would not be able to have a trustable 4D EFT.

\subsection{Non-perturbative and $\alpha'$ effects}
\label{secF}

The directions which are flat at leading order can be lifted by any effect which breaks the no-scale structure. These include $\alpha'$ corrections to the tree-level K\"ahler potential and non-perturbative contributions to the superpotential. When we study K\"ahler moduli stabilisation, we shall consider the $S$ and $U$-moduli fixed at their tree-level VEV which will be only slightly shifted by the subleading effects we are considering. 

The first $\alpha'$ correction to the effective action arise at $\mc{O}(\alpha'^3)$ and modifies the tree-level $K$ as follows (we are focusing only on the $T$-dependent part) \cite{Becker:2002nn}:
\be
K= -2\, {\rm ln} \,\left({\cal V} +\, \frac{\zeta}{2} \right)\qquad\text{with}\qquad \zeta = - \frac{\chi(X) \, \zeta(3)}{2 (2 \pi)^3\, g_s^{3/2}}\,.
\label{Ka}
\ee
Other perturbative corrections to $K$ arise at $\mc{O}(g_s^2 \alpha'^2)$ from Kaluza-Klein string loops and at $\mc{O}(g_s^2 \alpha'^4)$ from winding loops \cite{Berg:2005ja, Berg:2007wt}. At first sight, $g_s$ Kaluza-Klein loops might seem to be the dominant effect but, due to the extended no-scale cancellation, their contribution to the scalar potential arises effectively only at 2-loop $\mc{O}(g_s^4 \alpha'^4)$ level \cite{Cicoli:2007xp}. On the other hand $\mc{O}(g_s^2 \alpha'^4)$ winding loops are suppressed with respect to (\ref{Ka}) by both $g_s$ and $1/\vo$ factors. Finally the scalar potential get corrected also by higher derivative $F^4$ terms at $\mc{O}(\alpha'^3)$ but these terms are again $\vo$-suppressed with respect to (\ref{Ka}) \cite{Ciupke:2015msa}.

The tree-level superpotential receives instead non-perturbative corrections from the two ED3-instantons wrapping the two dP$_8$ cycles:
\be
W = W_0 + A_2\, e^{ - 2\pi T_2} + A_3\,e^{-2\pi T_3}\qquad\text{with}\qquad W_0 = \langle \int_X G_3\wedge \Omega \rangle\,.
\label{Wtot}
\ee
Plugging (\ref{Ka}) and (\ref{Wtot}) into the general expression for the $N=1$ supergravity F-term scalar potential, we obtain three contributions: 
\be
V_F = V_{\alpha'} + V_{\rm np1} + V_{\rm np2}\,,
\label{oddVgen1}
\ee
where (writing $W_0 = |W_0|\,e^{{\rm i}\theta_0}$, $A_2 = |A_2|\,e^{{\rm i}\theta_2}$ and $A_3 = |A_3|\,e^{{\rm i}\theta_3}$):
\bea
\label{Vtot1}
V_{\alpha'} &=& \, \frac{12 \, \zeta \,  |W_0|^2}{\left(2\vo+ \zeta \right)^2 \left(4\vo- \zeta \right)} \\
V_{\rm np1}  &=& \, \sum_{i=2}^3 \, \frac{8\, |W_0|\, |A_i| \, e^{-2\pi\, \tau_i} \, \cos\left(2\pi \rho_i +\theta_0-\theta_i\right)}{\left(2\vo+ \zeta \right) \left(4\vo - \zeta \right)} \left(8\pi\, \tau_i + \frac{3\, \zeta}{\left(2\vo+\zeta \right)} \right) \\
V_{\rm np2} &=& \,  \sum_{i=2}^3\, \frac{64 \pi^2\, \sqrt{\tau_i} \, |A_i|^2\, e^{-4\pi\, \tau_i}}{\sqrt{2} \left(2\vo + \zeta \right)} + \frac{4\, |A_i|^2 \, e^{- 4\pi\, \tau_i}}{\left(2\vo+ \zeta \right) \left(4\vo- \zeta \right)} \left(16\pi\, \tau_i \left(2\pi\tau_i+1\right) + \frac{3 \zeta}{\left(2\vo+\zeta \right)} \right )\nonumber \\
&+& \frac{8\, |A_2|\, |A_3| \, e^{-2\pi\left(\tau_2 + \tau_3\right)}}{\left(2\vo+ \zeta \right) \left(4\vo- \zeta \right)} 
\cos\left[2\pi\left( \rho_2 -\rho_3\right) + \theta_3 +\theta_0-\theta_2 \right] \nonumber \\
&\times& \left(8\pi \, \left(\tau_2 + \tau_3 + 4\pi\, \tau_2 \tau_3 \right)+ \frac{3\zeta}{\left(2\vo+\zeta \right)} \right) \,. 
\label{Vtot2}
\eea
In Sec.~\ref{secInfl} we will use the complete expressions (\ref{Vtot1}) - (\ref{Vtot2}) to perform a numerical study of the full inflationary dynamics of our model where the inflaton can be either of the two blow-up modes $\tau_2$ and $\tau_3$. However, in order to develop an analytical understanding of moduli stabilisation, we shall now approximate the scalar potential by considering only the leading order terms in the large volume limit $\vo\gg \zeta$. This leads to a typical LVS scalar potential of the form:
\be
V_\LVS = \sum_{i=2}^3 \left( \frac{32 \pi^2\, \sqrt{\tau_i} \, |A_i|^2\, e^{-4\pi\, \tau_i}}{\sqrt{2} \vo}+\frac{8\pi\, |W_0|\, |A_i| \,\tau_i\, e^{-2\pi\, \tau_i} \, \cos\left(2\pi \rho_i +\theta_0-\theta_i\right)}{\vo^2} \right) 
+ \frac{3 \, \zeta \,  |W_0|^2}{4\vo^3}\,.
\label{VLVS}
\ee
Notice that the cross term between the two blow-up modes drops out at leading order, and so the two axions $\rho_2$ and $\rho_3$ are fixed at:
\be
\rho_i = k+\frac12+\frac{(\theta_i-\theta_0)}{2\pi} \qquad\text{with}\quad k\in\mathbb{Z}\qquad\forall i=2,3\,.
\label{axMin}
\ee
The potential (\ref{VLVS}) has to be supplemented with the contributions from the soft scalar masses of the open string modes $\varphi$ and $C$ which read:
\be
V_{\rm soft} = m_\varphi^2 |\varphi|^2 + m_C^2 |C|^2\,,
\ee
where the generic expression of the scalar mass $m_0$ involves the gravitino mass $m_{3/2}$, the moduli F-terms and the K\"ahler metric for matter fields $\tilde{K}$ as follows: 
\be
m_0^2 = m_{3/2}^2 - F^I F^{\bar{I}}\partial_I\partial_{\bar{J}}\ln\tilde{K}\,.
\label{m0}
\ee
The K\"ahler metric for $\varphi$ depends just on the dilaton $S$ since it is given by $\tilde{K}_\varphi = 1/{\rm Re}(S)$. Given that $S$ is fixed supersymmetrically, i.e.\ $F^S=0$ (at least at leading order), the mass term for the hidden sector matter field $\varphi$ is simply given by the gravitino mass:
\be
m_\varphi^2 = m_{3/2}^2 = e^K |W|^2 = \frac{e^{K_{\rm cs}}\, |W_0|^2}{2\, {\rm Re}(S)\, \vo^2}\,.
\label{m32}
\ee
On the other hand, the K\"ahler metric for $C$ depends also on the $T$-moduli since $\tilde{K}_C = 1/\tau_1 +\cdots$ where the dots represent corrections beyond tree-level. Plugging this expression of $\tilde{K}_C$ into the general formula (\ref{m0}) for soft scalar masses, one finds a leading order cancellation between the gravitino mass and the non-zero F-terms of the `large' K\"ahler modulus $\tau_1$ which is due to the underlying no-scale structure \cite{Blumenhagen:2009gk, Aparicio:2014wxa}. Due to the locality of the visible sector which determines the form of $\tilde{K}_C$, the visible sector field $C$ acquires a mass of order $m_{3/2}/\sqrt{\vo}$ which is suppressed with respect to the gravitino mass \cite{Blumenhagen:2009gk, Aparicio:2014wxa}. Therefore we end up with:
\be
V_{\rm soft} = \frac{c_2\,m_{3/2}^2}{q_\varphi \vo^{2/3}} + m_C^2 |C|^2\,,
\label{Vsoft}
\ee
where we have written $\varphi$ in terms of $\vo$ according to the first D-term stabilisation condition in (\ref{Dfix}). If $m_C^2>0$, the visible sector field $C$ is fixed at zero size, i.e.\ $|C|=0$. From the second D-term condition in (\ref{Dfix}) this, in turn, implies $\tau_4=0$ ensuring that the dP$_0$ divisor supporting the visible sector is collapsed to zero size. This result is very robust since, due to the sequestering effect, $\tau_4$ remains in the singular regime, i.e.\ it develops a VEV below the string scale, even if $C$ develops a tachyonic mass \cite{Cicoli:2013cha}. 

Setting $|C|=0$, the total F-term potential therefore becomes (including the correct normalisation of $V_\LVS$):\footnote{The $\varphi$-dependence in (\ref{Vsoft}) gives rise to a shift of the first D-term relation in (\ref{Dfix}) which is $\vo$-suppressed, and so we shall neglect it. Moreover, \eqref{m32} guarantees that $\mathcal{C}_{\rm up}$ is positive.}
\be
V_{\rm tot} =  \frac{e^{K_{\rm cs}}}{2\,{\rm Re}(S)} \left( V_\LVS + \frac{\cC_{\rm up}\, |W_0|^2}{\vo^{8/3}} \right)
\qquad\text{with}\qquad \cC_{\rm up} =  \frac{c_2}{q_\varphi} = \frac{I^A_{U(4)}}{24\pi\,q_\varphi} \left(\frac23\right)^{1/3} >0\,.
\label{Vtot}
\ee
In the limit where $\epsilon_i = \frac{1}{8\pi\tau_i}\ll 1$, the global minimum of the total potential (\ref{Vtot}) is located at:
\bea
\label{Tstab1}
\vo &=& \frac{\sqrt{2} \, \sqrt{\tau_i}\,(1-4\epsilon_i)}{8\pi  (1 - \epsilon_i)}\,\frac{|W_0|}{|A_i|}  \,e^{2\pi \tau_i}
\simeq \frac{\sqrt{2} \, \sqrt{\tau_i}}{8\pi}\,\frac{|W_0|}{|A_i|}  \,e^{2\pi \tau_i} \qquad \forall i =2,3 \,, \\
\frac{3\,\zeta}{2\sqrt{2}}  &=& \sum_{i=2}^3 \frac{(1-4\epsilon_i)}{(1-\epsilon_i)^2}\,\tau_i^{3/2}-\frac{8\sqrt{2}\,\cC_{\rm up}}{27}\, \vo^{1/3}
\simeq \tau_2^{3/2}+\tau_3^{3/2}-\frac{8\sqrt{2}\,\cC_{\rm up}}{27}\,\vo^{1/3}\,.
\label{Tstab2}
\eea
We point out that (\ref{Tstab1}) implies:
\be
e^{2\pi (\tau_2-\tau_3)}  \simeq \sqrt{\frac{\tau_3}{\tau_2}} \frac{|A_2|}{|A_3|}\,,
\label{simple}
\ee
and so the difference between the two blow-up modes is controlled by the two prefactors of the non-perturbative effects $|A_2|$ and $|A_3|$. If all moduli are fixed at their minimum, the resulting vacuum energy turns out to be (neglecting $\mc{O}(\epsilon)$ corrections): 
\be
\langle V_{\rm tot} \rangle \simeq \, \frac{e^{K_{\rm cs}} |W_0|^2}{18\,{\rm Re}(S)\, \vo^3} 
\left[\cC_{\rm up} \vo^{1/3} - \frac{9 \sqrt{2}}{4\pi}\left(\sqrt{\tau_2}+\sqrt{\tau_3}\right) \right]\,.
\label{CC}
\ee
Notice that this potential scales as $\mc{O}(1/\vo^3)$, and so non-perturbative and $\alpha'$ effects are indeed subdominant with respect to background fluxes and D-terms. The vacuum energy (\ref{CC}) can be zero (or slightly positive to get a dS vacuum) if the gauge and background fluxes are tuned so that:
\be
\cC_{\rm up} \vo^{1/3} = \frac{9 \sqrt{2}}{8\pi}\left(\sqrt{\tau_2}+\sqrt{\tau_3}\right)\,.
\label{CCtuning}
\ee
Plugging this result back in (\ref{Tstab2}) we obtain:
\be
\frac{3\,\zeta}{2\sqrt{2}} = \sum_{i=2}^3 \tau_i^{3/2}\left(1-18\epsilon_i\right) \simeq  \tau_2^{3/2}+\tau_3^{3/2}\,,
\label{Tstab2new}
\ee
showing that the sum of the two dP$_8$ divisors is controlled just by the $\alpha'$ parameter $\zeta$ which depends on the Calabi-Yau Euler number $\chi(X)$ and the string coupling $g_s$.

\subsection{Moduli mass spectrum and soft terms}
\label{SUSY}

Before presenting some explicit choices of the underlying parameters which give rise to a dS vacuum with all closed string moduli stabilised, let us describe what is the resulting moduli mass spectrum. The closed string moduli fixed at $\mc{O}(1/\vo^2)$ are the dilaton $S$, the complex structure moduli $U_\alpha$, $\alpha=1,...,h^{1,2}_-$, and the K\"ahler modulus $T_4=\tau_4+{\rm i}\rho_4$. The volume of the dP$_0$ divisor $\tau_4$ acquires a mass of order the string scale while the associated axion $\rho_4$ is eaten up by the anomalous $U(1)$ at the singularity in the process of anomaly cancellation. On the other hand, $S$ and the $U$-moduli develop a mass of order $m_{3/2}$. At this level of approximation the vacuum is Minkowski and supersymmetry is broken along the three K\"ahler moduli $T_1$, $T_2$ and $T_3$ which are however still flat directions.

These directions are lifted at $\mc{O}(1/\vo^3)$ by subdominant non-perturbative and $\alpha'$ effects. The two blow-up modes $\tau_2$ and $\tau_3$, and their associated axions $\rho_2$ and $\rho_3$, develop a mass of order $m_{3/2}$. Let us stress that the dilaton and the complex structure moduli can be safely integrated out even if their mass is of the same order of magnitude since they are decoupled (at leading order) from the K\"ahler moduli as can be seen from the factorised form of the tree-level K\"ahler potential in (\ref{Ktree}). The divisor volume $\tau_1$ which controls the overall volume acquires instead a lower mass of order $m_{3/2}/\sqrt{\vo}$. Given that this modulus is fixed via perturbative effects, its axionic partner $\rho_1$, remains still massless at this level of approximation. This axionic direction gets lifted by $T_1$-dependent non-perturbative corrections to the superpotential. The resulting mass for $\rho_1$ is exponentially suppressed with respect to the gravitino mass since it scales as $m_{3/2}\,e^{-\vo^{2/3}}$.

In the final dS vacuum supersymmetry is broken mainly along the K\"ahler moduli directions which develop non-vanishing F-terms of order $F^{T_i}/\tau_i \sim m_{3/2}$ $\forall i=1,2,3$. The dilaton and the complex structure moduli also develop non-zero F-terms since their tree-level supersymmetric VEVs are shifted by non-perturbative and $\alpha'$ effects. However these F-terms are subdominant since they scale as $F^S\sim F^U \sim m_{3/2}/\vo$. The F-term of the dP$_0$ modulus $T_4$ remains instead zero. 

Supersymmetry breaking is mediated to the visible sector at the dP$_0$ singularity via gravitational interactions. However, since the visible sector is localised at a singularity, sequestering effects give rise to soft terms which are suppressed with respect to the gravitino mass.\footnote{Note however that for D3-branes at orientifold singularities threshold corrections to the gauge kinetic function might induce moduli redefinitions which could spoil sequestering \cite{Conlon:2010ji}.} In particular, squark and slepton masses scale as $m_0^2\sim m_{3/2}^2/\vo$ while gaugino masses arise only at subleading order (since they are generated by the F-term of the dilaton) and scale as $M_{1/2}^2\sim m_{3/2}^2/\vo^2$ \cite{Blumenhagen:2009gk, Aparicio:2014wxa}.\footnote{In models with T-brane dS uplifting, i.e. where dS vacua are obtained from non-zero F-terms of hidden matter fields, scalar masses are always hierarchically larger than gaugino masses due to non-vanishing D-terms from the hidden uplifting sector (barring unexpected cancellations) \cite{Aparicio:2014wxa}.} The $\mu$-term, which sets the Higgsino mass, is also of order $M_{1/2}$ (if it is generated from K\"ahler potential contributions). This leads to a split supersymmetry scenario with TeV-scale gauginos and neutralinos for values of the volume of order $\vo\sim 10^6 - 10^7$ \cite{Blumenhagen:2009gk, Aparicio:2014wxa}.

\subsection{Choices of underlying parameters}
\label{dSnumerics}

In this section we shall present some choices of the underlying parameters which allow for an explicit stabilisation of all K\"ahler moduli in a Minkowski (or slightly dS) vacuum. As can be seen from (\ref{phicharge}), the $U(1)$ charge of the hidden sector field $\varphi$ is $q_\varphi=2$. Moreover, the integer $n$ which fixes the gauge flux on the D7-stack (and, in turn, also the $U(1)$ charge of the closed string modulus $T_1$) can consistently take only the values listed in (\ref{possibn}). We will also take into account $N=1$ corrections to the Calabi-Yau Euler number due to the presence of O7-planes. This lead to an `effective' Euler characteristic defined as \cite{Minasian:2015bxa}:
\be
\chi_{\rm eff} = \chi(X) + 2\, \int_X D_{O7}^3 =-420 + 2 \, \left( \int_X \, D_Y^3 + \int_X \, D_Z^3 \right)=-24\,,
\label{eq:chi_eff}
\ee
where we have used the fact in our case the O7-planes consist of two non-intersecting components $Z=0$ and $Y=0$.\footnote{Strictly speaking, the correction of \cite{Minasian:2015bxa} has been computed for a configuration with one O7-plane and one fully recombined invariant D7-brane that cancel the D7-tadpole. We assume here that such a correction persist in the form (\ref{eq:chi_eff}) also for our different configuration.} 

\begin{figure}[H]
\begin{center}
\hspace*{-1cm} \includegraphics[width=15.0cm,height=10.5cm]{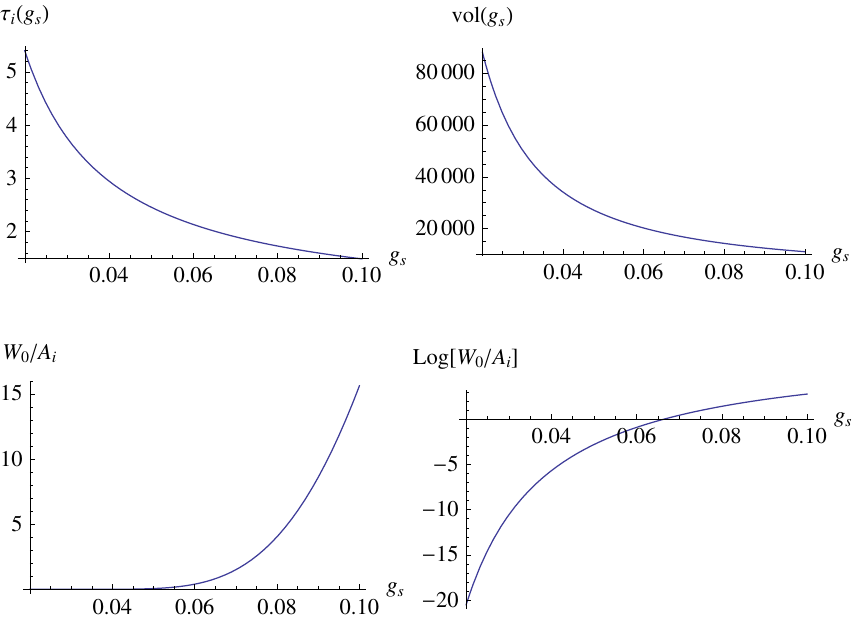}
\caption{Values of $\tau_s$, $\vo$ and $|W_0|/|A_s|$ which give $\langle V\rangle=0$ as a function of the string coupling $g_s$.} 
\label{plotall}
\end{center}
\end{figure}

Thus the parameters which we can choose (allowing an appropriate tuning of the background fluxes) are five: $n$, $g_s$, $|W_0|$, $|A_2|$ and $|A_3|$. We shall now present some illustrative choices of these five parameters which lead to a Minkowski vacuum in the simple situation where $|A_2|=|A_3|\equiv |A_s|$. As explained around (\ref{simple}), this simplification forces the two dP$_8$ moduli to have the same VEV: $\tau_2=\tau_3 \equiv \tau_s$. Moreover we will set the flux parameter $n=-1$ (which leads to $\cC_{\rm up} = (3/2)^{5/3}/(4\pi) = 0.1564$), and so we end up with only three free parameters $g_s$, $|W_0|$ and $|A_s|$. The condition for a vanishing cosmological constant (\ref{CCtuning}) fixes just one of them while the other two remain free and can be used to obtain TeV-scale gauginos and to ensure that the string coupling is in the perturbative regime. Tab.~\ref{TabdS} presents five different choices of these parameters for $g_s\ll 1$ while Fig.~\ref{plotall} shows more in general how $\tau_s$, $\vo$ and the ratio $|W_0|/|A_s|$ vary with the string coupling $g_s$.

\begin {table}[H]
\begin{center}
 \begin{tabular}{|c || c | c | c || c | c | c |} 
 \hline
 & & &  & & & \\
$g_s$ & $|W_0|/|A_s|$ & $\langle\tau_s\rangle$ & $\langle\vo\rangle$ & $|W_0|/|A_s|$ & $\langle\tau_s\rangle$ &$\langle\vo\rangle$ \\ 
 & & & & & &\\
 \hline\hline
 0.10 & 15.627 & 1.486 & 11118.88 & 17.776  & 1.437 & 9166.91 \\ 
 \hline
 0.08 & 4.039 & 1.727 & 14318.12 & 4.053  & 1.726 & 14262.42 \\
 \hline
 0.06 & 0.3973 & 2.132 & 20250.11 & 0.3968 & 2.133 & 20289.57 \\
 \hline
 0.04 & $3.345 \times 10^{-3}$ & 2.947 & 34064.05 & $3.342 \times 10^{-3}$  & 2.947 & 34131.09 \\
  \hline
 0.02 & $1.264 \times 10^{-9}$ & 5.400 & 87919.95 & $1.262 \times 10^{-9}$ & 5.401 & 88091.51 \\
 \hline
\end{tabular}
\caption{Five choices of the underlying parameters which lead to a Minkowski vacuum. The values on the left have been obtained by using the leading order potential while the values listed on the right have been obtained using the full scalar potential. Notice that the leading order potential produces the global minimum with quite good accuracy for small string coupling.} 
\label{TabdS}
\end{center}
\end {table}

\section{Inflation}
\label{secInfl}

The form of the Calabi-Yau volume (\ref{CYvo}) and the scalar potential (\ref{VLVS}) are particularly suitable to realise an inflationary model where the inflaton is the blow-up mode $\tau_2$ \cite{Conlon:2005jm}. If this field is displaced from its minimum, it experiences a very flat potential due to the exponential suppression coming from the $T_2$-dependent non-perturbative effects. If during the inflationary evolution both $\tau_1$ and $\tau_3$ are kept at their minimum, the single-field inflationary potential takes the simple form:\footnote{As pointed out in \cite{Cicoli:2008gp}, any kind of $\tau_2$-dependent perturbative correction to $K$ \cite{Berg:2005ja, Berg:2007wt, Cicoli:2007xp, Ciupke:2015msa} could cause an $\eta$-problem and spoil the flatness of the inflationary plateau. In what follows, we shall therefore assume that the coefficients of these dangerous perturbative corrections can be suitably tuned to avoid any $\eta$-problem.}
\be
V_{\rm inf} \simeq V_0 - \frac{8\pi\, |W_0|\, |A_2| \,\tau_2\, e^{-2\pi\, \tau_2}}{\vo^2} \,,
\label{VLVSinf}
\ee
where we have neglected the $T_2$-dependent non-perturbative term with a double suppression. Interestingly, the stabilised values of the moduli presented in Tab.~\ref{TabdS} are in the right ballpark to reproduce the correct amplitude of the density perturbations. 

\subsection{The need for a multi-field analysis}

In order to decouple the volume mode from the inflationary dynamics or, in other words, to keep it fixed during inflation, the authors of \cite{Conlon:2005jm} assumed the presence of $n\gg 1$ blow-up modes. In fact, in this case the minimisation condition (\ref{Tstab2new}) would be modified to (assuming that each `small' divisor is wrapped by an ED3-instanton):
\be
\frac{3\,\zeta}{2\sqrt{2}} = \sum_{i=2}^n \tau_i^{3/2}\,.
\label{Tstab2n}
\ee
When $\tau_2$ is displaced from its minimum, the $\tau_2$ dependent terms in the scalar potential are exponentially suppressed with respect to the rest, and so can be safely neglected. Hence the minimisation relation (\ref{Tstab2n}) simplifies to:
\be
\frac{3\,\zeta}{2\sqrt{2}} = \sum_{i=3}^n \tilde{\tau}_i^{3/2}\,,
\label{Tstab2ne}
\ee
where $\tilde{\tau}_i$ is the new value of the $i$-th blow-up mode. If $\tilde{\tau}_i\neq \tau_i$, the relations (\ref{Tstab1}) would cause the volume to get destabilised from its initial VEV. The stability requirement $\tilde{\tau}_i\simeq \tau_i$ therefore translates into:
\be
\sum_{i=2}^n \tau_i^{3/2} = \sum_{i=3}^n \tau_i^{3/2} \left(1+\frac{\tau_2^{3/2}}{\sum_{i=3}^n \tau_i^{3/2}} \right) \simeq \sum_{i=3}^n \tau_i^{3/2}
\quad\Leftrightarrow\quad \frac{\tau_2^{3/2}}{\sum_{i=3}^n \tau_i^{3/2}} \ll 1\,,
\label{stab}
\ee
which can be easily satisfied for $n\gg 1$. In our case however $n=3$, and so the stability condition (\ref{stab}) reduces to $\tau_2\ll \tau_3$. From (\ref{simple}), we can clearly see that this condition can be satisfied only if $|A_2|\ll |A_3|$. Even if the two pre-factors of the non-perturbative effects are tuned to achieve the required hierarchy,\footnote{$|A_1|$ and $|A_2|$ are tunable as they are functions of the flux dependent complex structure moduli.} it is still necessary to go beyond the single-field dynamics described by the potential (\ref{VLVSinf}) to study the full three-field inflationary evolution since, in the case with only two blow-up modes, the volume shift during inflation can never be completely ignored. However let us point out that the single-field potential (\ref{VLVSinf}) still provides a good qualitative understanding of the reason why we can obtain a potential which is flat enough to drive inflation even in the more general multi-field case.

\subsection{Multi-field inflationary evolution}

In this section we shall follow ref. \cite{BlancoPillado:2009nw} and perform a numerical multi-field analysis to find inflationary trajectories which reproduce enough efoldings of inflation and are stable throughout all inflationary dynamics, from given initial conditions to the final settling of the fields into the global Minkowski minimum. We shall satisfy the stability condition (\ref{stab}) by choosing $|A_2|$ hierarchical smaller than $|A_3|$, so that $\tau_2\ll \tau_3$ with still $2\pi\tau_2$ slightly larger than unity in order to be able to neglect higher order instanton contributions to the superpotential.

The three-field evolution is governed by the following Einstein-Friedmann equations:
\bea
\label{Friedmann}
\frac{d^2 \phi^a}{dN^2}&+&{\Gamma^a}_{bc}\frac{d\phi^b}{dN}\frac{d\phi^c}{dN}+\left(3+\frac{1}{H}\frac{dH}{dN}\right)\frac{d\phi^a}{dN}+\frac{g^{ab} \partial_b V}{H^2}=0\,, \\
3\,H^2&=& V(\phi^a)+\frac12 H^2\, g_{ab} \frac{d\phi^a}{dN}\frac{d\phi^b}{dN} \,,
\label{constran}
\eea
where $g_{ab}$ is the field space metric, $\Gamma^a_{bc}$ are the associated Christoffel symbols and $N$ is the number of efoldings which we are using as the time coordinate during the evolution via $dN = H dt$. Using (\ref{Friedmann}) and (\ref{constran}), the variation of the Hubble rate in terms of the number of efoldings can be expressed as: 
\be
\frac{1}{H}\frac{dH}{dN}=\frac{V}{H^2}-3\,.
\label{third}
\ee
Thus the generic expression of the slow-roll parameter $\epsilon$ takes the form:
\be
\epsilon \equiv - \frac{1}{H^2} \frac{dH}{dt} =  \frac{1}{H} \frac{dH}{dN} = \frac12\, g_{ab}\, \frac{d\phi^a}{dN} \, \frac{d\phi^b}{dN}\,.
\label{eq:epsilon-gen}
\ee
Notice that this definition of $\epsilon$ holds beyond the single field as well as the slow-roll approximation. In the slow-roll regime, (\ref{eq:epsilon-gen}) simplifies to:
\be
\epsilon_s = \frac{g^{ab}\, \partial_a V \, \partial_b V}{2\, V^2}\,.
\label{eq:epsilon_slow-roll}
\ee
The power spectrum and spectral index for scalar perturbations are given by:
\be
P_s = \frac{1}{150\, \pi^2}\, \frac{V}{\epsilon}\,, \qquad n_s = 1 + \frac{d\, \ln(P_s(N))}{dN}\,,
\ee
where the COBE normalisation for the amplitude is $P_s = 3.7 \times 10^{-10}$. 

We shall now solve the evolution equations (\ref{Friedmann}) and (\ref{constran}) numerically, considering the complete expressions (\ref{Vtot1}) - (\ref{Vtot2}) for the scalar potential with the axions fixed at their global minimum (\ref{axMin}). We shall also include an uplifting term with subleading corrections of the form:
\be
V_{\rm up} = \frac{e^{K_{\rm cs}}}{2\,{\rm Re}(S)} \frac{\cC_{\rm up}\, |W_0|^2}{\vo^{8/3}} \left(1- \frac{\cC_{\rm sub}}{\vo^{2/3}} \right)\,,
\label{Vup}
\ee
where:
\be
\cC_{\rm sub} = \frac{e^{K_{\rm cs}} |W_0|^2}{4{\rm Re}(S)\, c_1\,c_2\,q_\varphi}
=\frac{e^{K_{\rm cs}} |W_0|^2}{{\rm Re}(S)}\left(\frac32\right)^{2/3}\,.
\ee

\subsection{Numerical analysis}

We shall now set the gauge fluxes as in Sec.~\ref{dSnumerics}, i.e.\ $n=-1$, and perform appropriate choices of the remaining four free parameters, $g_s$, $|W_0
|$, $|A_2|$ and $|A_3|$ which allow for a global Minkowski minimum and a viable inflationary dynamics. As studied in \cite{Cicoli:2016olq}, after the end of inflation the volume modulus drives an epoch of matter dominance which reduces the number of efoldings to $N_e \simeq 45$. We shall therefore evaluate the two main cosmological observables, the scalar spectral index $n_s$ and the tensor-to-scalar ratio $r$, at horizon exit which in this model takes place around $45$ efoldings before the end of inflation. At this point in field space we shall also make sure that the inflationary potential reproduces the correct amplitude of the density perturbations, taking into account the correct normalisation of the scalar potential by a factor of $g_s\, e^{K_{\rm cs}}/{(8\, \pi)}$ (see App. A of \cite{Burgess:2010bz}). 

Before presenting the results of our numerical analysis, let us mention that the period of matter domination due to the light volume mode leads to a very low reheating temperature of order $T_{\rm rh}\simeq \mc{O}(1-10)$ GeV with important implications for non-thermal WIMP dark matter \cite{Allahverdi:2013noa, Allahverdi:2014ppa, Aparicio:2015sda, Aparicio:2016qqb}, axionic dark radiation \cite{Cicoli:2012aq, Higaki:2012ar} and Affleck-Dine baryogenesis \cite{Allahverdi:2016yws}. In particular, the volume axion $\rho_1$ is ultra-light (since it acquires mass only via subleading non-perturbative effects suppressed by $e^{-2\pi\vo^{2/3}}\ll 1$), and so it behaves as an extra relativistic degree of freedom which contributes to $N_{\rm eff}$ \cite{Cicoli:2012aq, Higaki:2012ar}. Due to the non-vanishing branching ratio for the decay of $\tau_1$ into $\rho_1$, the ultra-light axion $\rho_1$ is produced at reheating, leading generically to $\Delta N_{\rm eff}\simeq \mc{O}(0.5-1)$ (depending on the strength of the volume mode coupling to Higgses and the ratio between its mass and scalar masses) \cite{Cicoli:2015bpq}. In the comparison of our model with cosmological observations, this extra amount of dark radiation should be imposed as a prior for Planck data analysis. If this is done, the value of the spectral index becomes closer to unity. In fact, the Planck paper \cite{Ade:2015xua}, presents an example with $\Delta N_{\rm eff}=0.39$ that gives $n_s= 0.983\pm 0.006$ (TT+lowP). In the following, we shall therefore look for parameter values which yield $n_s\simeq 0.98$ at $N_e\simeq 45$ efoldings before the end of inflation. 

The strategy used to find working values of the underlying parameters is the following: we considered a value of the string coupling which is still in the perturbative regime and then we found the range of parameters that lead to a stabilised value of the inflaton $\tau_2$ such that $2\pi\, \langle\tau_2\rangle$ is just slightly more than $1$ so that higher instanton effects can still be negligible. In this way the shift of $\tau_2$ from the minimum to drive inflation produces only a negligible shift of the local minimum of the other two moduli $\tau_1$ and $\tau_3$. Moreover, we focus on these initial conditions:
\be
\phi^a_{\rm in} = \left\{\tau_1^{\rm in}, \tau_2^{\rm in}, \tau_3^{\rm in} \right\}\, , \qquad  \left.\left(\frac{d\phi^a}{dN}\right)\right|_{\phi^a = \phi^a_{\rm in}} = \{0, 0, 0\}\qquad \forall \, i = 1, 2, 3\,.
\ee
Notice that even if we start with zero initial velocity for each field, the actual values for these field derivative variations are attained within a few efoldings during the evolution. Tab.~\ref{tab5} shows the values of $g_s$, $|W_0|$, $|A_2|$ and $|A_3|$ and the corresponding VEVs of all K\"ahler moduli which lead to a Minkowski vacuum and a viable inflationary dynamics with the correct COBE normalisation and enough efoldings of inflation. On the other hand, Tab.~\ref{tab6} gives the initial conditions, the number of efoldings, $n_s$, $r$ and the VEV of the tree-level complex structure K\"ahler potential needed to obtain $P_s = 3.7\times 10^{-10}$. 

\begin {table}[H]
\begin{center}
 \begin{tabular}{|c |c| c| c ||c |c | c | c |} 
 \hline
 & & & & & & &\\
$g_s$ & $|W_0|$ & $|A_2|^{-1}$ & $|A_3|^{-1}$ & $\langle\tau_1\rangle$ & $\langle\tau_2\rangle$ & $\langle\tau_3\rangle$ & $\langle\vo\rangle$ \\ 
 & & & & & & &\\
 \hline\hline
 0.25 & 2.70937 & $2.0 \times 10^5$ & $5$ & 313.389 & 0.162328 & 1.12956 & 871.165 \\
 \hline
 0.20 & 0.577542 & $4.5 \times 10^6$ & $10$ & 367.954 & 0.160021 & 1.29046 & 1108.36 \\
 \hline
 0.15 & 0.119824 & $1.0 \times 10^7$ & $12$ & 457.946 & 0.161688 & 1.54674 & 1538.97 \\
 \hline
 0.10 & 0.00780641 & $1.9 \times10^8$ & $10$ & 630.098 & 0.162440 & 2.06034 & 2483.91 \\
 \hline
 0.05 & $8.2095 \times10^{-7}$ & $1.8 \times10^{12}$ & $10$ &1166.73 & 0.167371 & 3.61622 & 6258.91 \\
 \hline
\end{tabular}
\caption{Five choices of parameters $g_s$, $|W_0|$, $|A_2|$ and $|A_3|$ and corresponding stabilised values of all K\"ahler moduli which reproduce a Minkowski minimum and a viable inflationary dynamics.} 
\label{tab5}
\end{center}
\end{table}

\begin {table}[H]
\begin{center}
 \begin{tabular}{|c |c| c| c ||c |c | c | c || c|} 
 \hline
 & & & & & & & &\\
$\tau_1^{\rm in}$ & $\tau_2^{\rm in}$ & ${\tau}_3^{\rm in}$ & $\vo^{\rm in}$ &  $N_e$ & $r$  & $n_s$ & $P_s (K_{\rm cs} = 0)$ & ${K_{\rm cs}}$ \\ 
 & & & & & & & &\\
 \hline\hline
 356.395 & 3.13 & 1.15744 & 1054.03 & 46.4 &  $1.2\times 10^{-8}$ & 0.976  & $2.1\times 10^{-7}$  &  -6.33 \\ 
 \hline
 375.478 & 3.18 & 1.29461 & 1139.90 & 47.1 &  $1.0\times 10^{-8}$ & 0.981  & $1.3\times 10^{-9}$  &  -1.22 \\ 
 \hline
 485.000 & 3.23 & 1.55942 & 1674.70 & 45.4 &  $7.5\times 10^{-9}$ & 0.979  & $5.9\times 10^{-11}$  &  1.84 \\ 
 \hline
 669.930 & 3.32 & 2.07421 & 2720.43 & 47.2 &  $4.2\times 10^{-9}$ & 0.980  & $9.2\times 10^{-14}$  &  8.30 \\ 
 \hline
 1387.26 & 3.47 & 3.65655 & 8112.76 & 47.2 &  $1.4\times 10^{-9}$ & 0.973  & $2.2\times 10^{-22}$  &  28.15 \\ 
 \hline
\end{tabular}
\caption{Initial conditions, number of efoldings and predictions for the cosmological observables for the five parameter choices given in Tab.~\ref{tab5}. The values of $K_{\rm cs}$ are those needed for obtaining the correct COBE normalisation of the scalar power spectrum $P_s$ ($K_{\rm cs}$ can be positive or negative depending on the stabilised value of the complex structure moduli).} 
\label{tab6}
\end{center}
\end{table}

Let us make some comments on the values presented in Tab.~\ref{tab5} and \ref{tab6}:
\begin{itemize}
\item The VEV of $\tau_2$ in Tab.~\ref{tab5} is smaller than $1$. This is however the Einstein frame value of the volume of the corresponding dP$_8$ divisor in units of $\ell_s=2\pi\sqrt{\alpha'}$. This modulus is related to the corresponding string frame value as $\tau_{{\scriptscriptstyle E}} = \tau_{\rm st}/g_s$. Thus a 4D effective field theory analysis is under control if:
\be
\frac{\sqrt{\alpha'}}{\tau_{\rm st}^{1/4}} \ll 1 \qquad \Leftrightarrow \qquad \frac{1}{(2\pi) {(\tau_{{\scriptscriptstyle E}}\, g_s)}^{1/4}}\ll 1\,.
\ee
For all the parameter choices of Tab.~\ref{tab5}, this ratio is consistently smaller than unity.

\item Only the inflaton $\tau_2$ is shifted significantly from its global minimum while the remaining two fields do not move much during inflation. 

\item Decreasing $g_s$ increases $\zeta$, and so the blow-up mode $\tau_3$ becomes larger and the model turns out to be more stable.

\item Larger values of $\tau_3$ give a larger overall volume $\vo$. However, the particular form of the uplifting contribution (\ref{Vup}) sets an upper bound on $\vo$ in order to avoid a run-away in the volume direction. This is consistently achieved by reducing $|W_0|$ while increasing $\tau_3$ (i.e. decreasing $g_s$). 

\item $|A_2|$ is chosen in order to keep $2\pi\langle\tau_2\rangle$ slightly above unity so that $\vo$ and $\tau_3$ do not shift significantly during inflation. Thus, as can be seen qualitatively from (\ref{simple}), larger values of $\tau_3$ imply smaller, and so more tuned, values of $|A_2|$. 

\item In our model both of the blow-up modes are dP$_8$ divisors with the same intersection numbers. Moreover, both of them are wrapped by an ED3 instanton, resulting in the same coefficient in the exponent of the non-perturbative effects. The fact that these model-dependent parameters are the same for both $\tau_2$ and $\tau_3$ causes a small value of $\langle\tau_2\rangle$ and a tuned $|A_2|$. More general cases with different intersection numbers and different origins of the non-perturbative effects, can give rise to larger values of $\langle\tau_2\rangle$ and more natural values of $|A_2|$ closer to unity. 

\item As can be seen qualitatively from the single-field potential (\ref{VLVSinf}), larger values of $\vo$ yield a larger suppression of the density perturbations. Hence a correct COBE normalisation requires larger, and so more tuned, values of the tree-level complex structure K\"ahler potential $K_{\rm cs}$ which appears in the prefactor of the inflationary potential as $e^{K_{\rm cs}}$.

\item As explained in Sec.~\ref{SUSY}, TeV-scale gaugino masses require values of the volume of order $10^6$-$10^7$ while in our case the requirement to reproduce the correct amplitude of the density perturbations fixes $\vo\simeq 10^3-10^4$ and $|W_0|$ as in Tab. \ref{tab5}. Hence our model cannot lead to low-energy supersymmetry since the gravitino mass is of order $10^{13}$ GeV for all the parameter choices of Tab. \ref{tab5} while gaugino masses tend to lie around $10^{10}$ GeV.\footnote{Small values of $|W_0|$ are compensated by large values of $e^{K_{\rm cs}/2}$, so that $m_{3/2}$ is of the same order for all parameter choices of Tab. \ref{tab5}.} The main phenomenological implication of this result is that dark matter cannot be a standard WIMP (either thermally or non-thermally produced) but it should have a different origin. In LVS string compactifications a natural dark matter candidate is the light bulk axion $\rho_1$ \cite{Cicoli:2012sz}.

\item Our numerical results are based on the assumption that the full $\alpha'^3$ correction to the K\"ahler potential is captured by (\ref{Ka}) with the modification (\ref{eq:chi_eff}). However, a change in $\chi_{\rm eff}$ would only modify the numbers but not the possibility of getting inflation in an analogous setup.
\end{itemize}

\begin{figure}[H]
\begin{center}
\includegraphics[width=13.5cm,height=8.8cm]{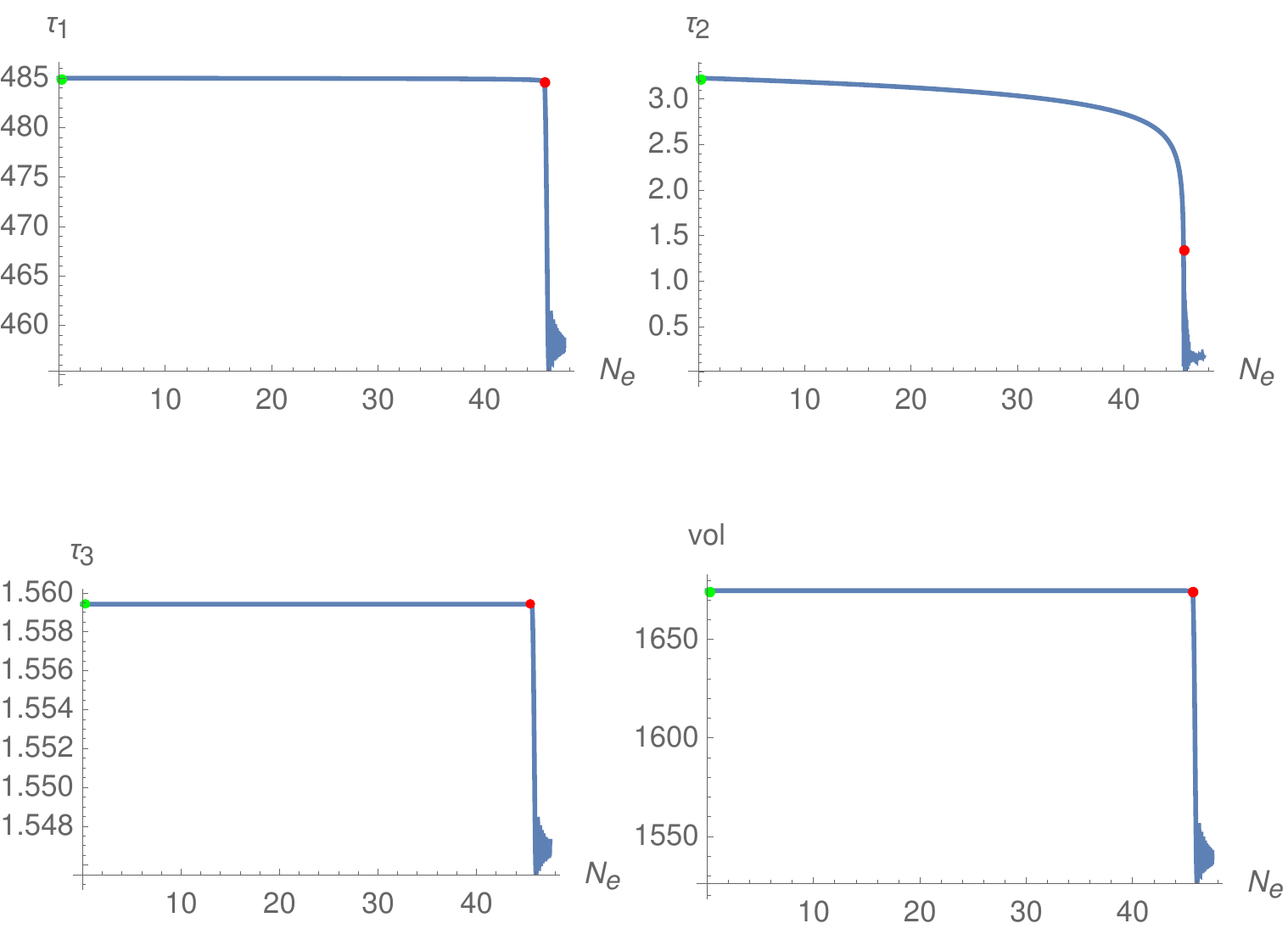}
\caption{Inflationary evolution of all fields during the whole inflationary epoch for the third case in Tab.~\ref{tab5} and \ref{tab6}. The green dot represents horizon exit while the red dot denotes the end of slow-roll inflation where $\epsilon_s=1$.} 
\label{Plots1}
\end{center}
\end{figure}

\begin{figure}[H]
\begin{center}
\includegraphics[width=13.5cm,height=8.8cm]{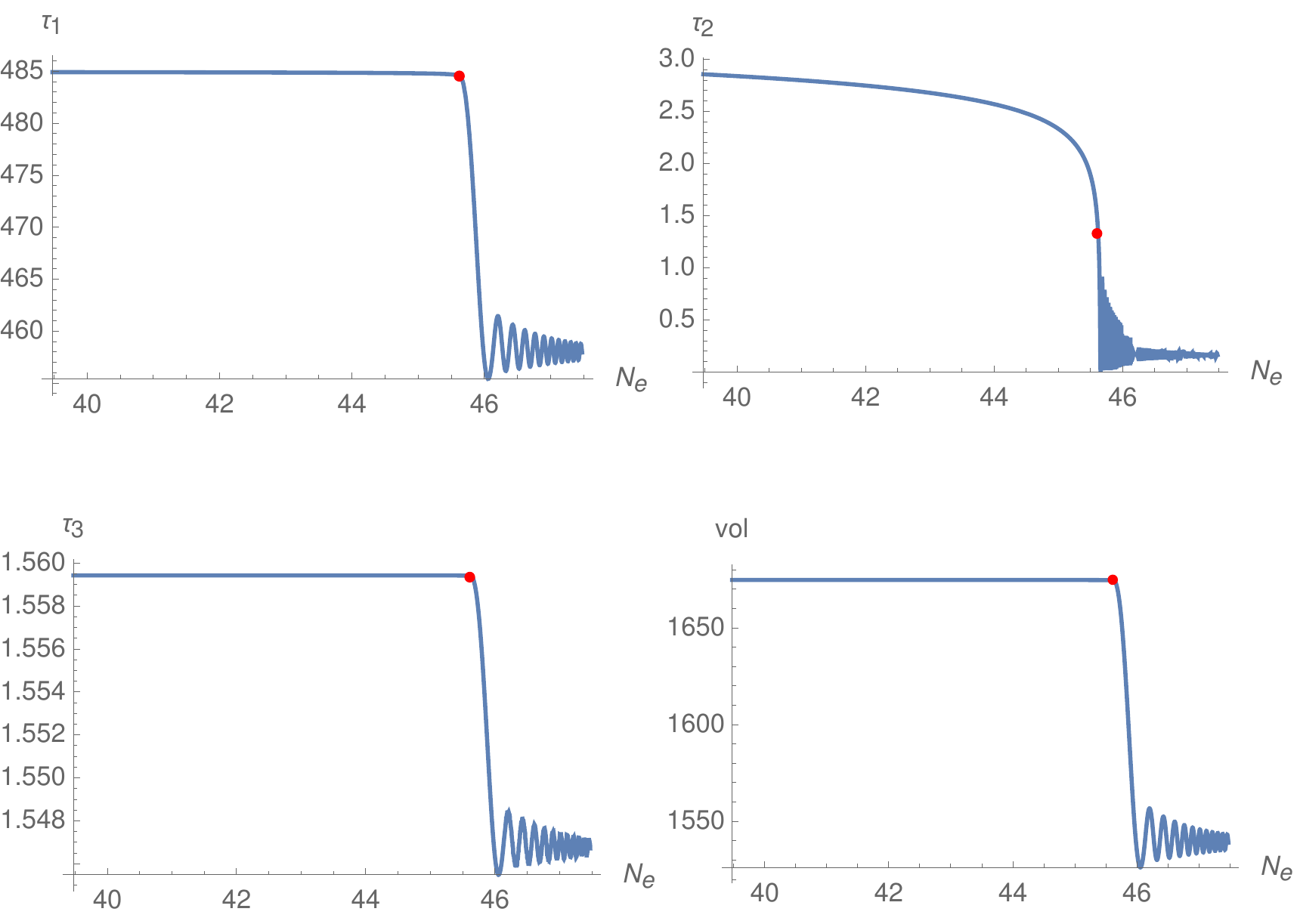}
\caption{Inflationary evolution of all fields during the last $8$ efoldings for the third case in Tab.~\ref{tab5} and \ref{tab6}. The red dot denotes the end of slow-roll inflation where $\epsilon_s=1$.} 
\label{Plots2}
\end{center}
\end{figure}

\begin{figure}[H]
\begin{center}
\includegraphics[width=14.0cm,height=8.8cm]{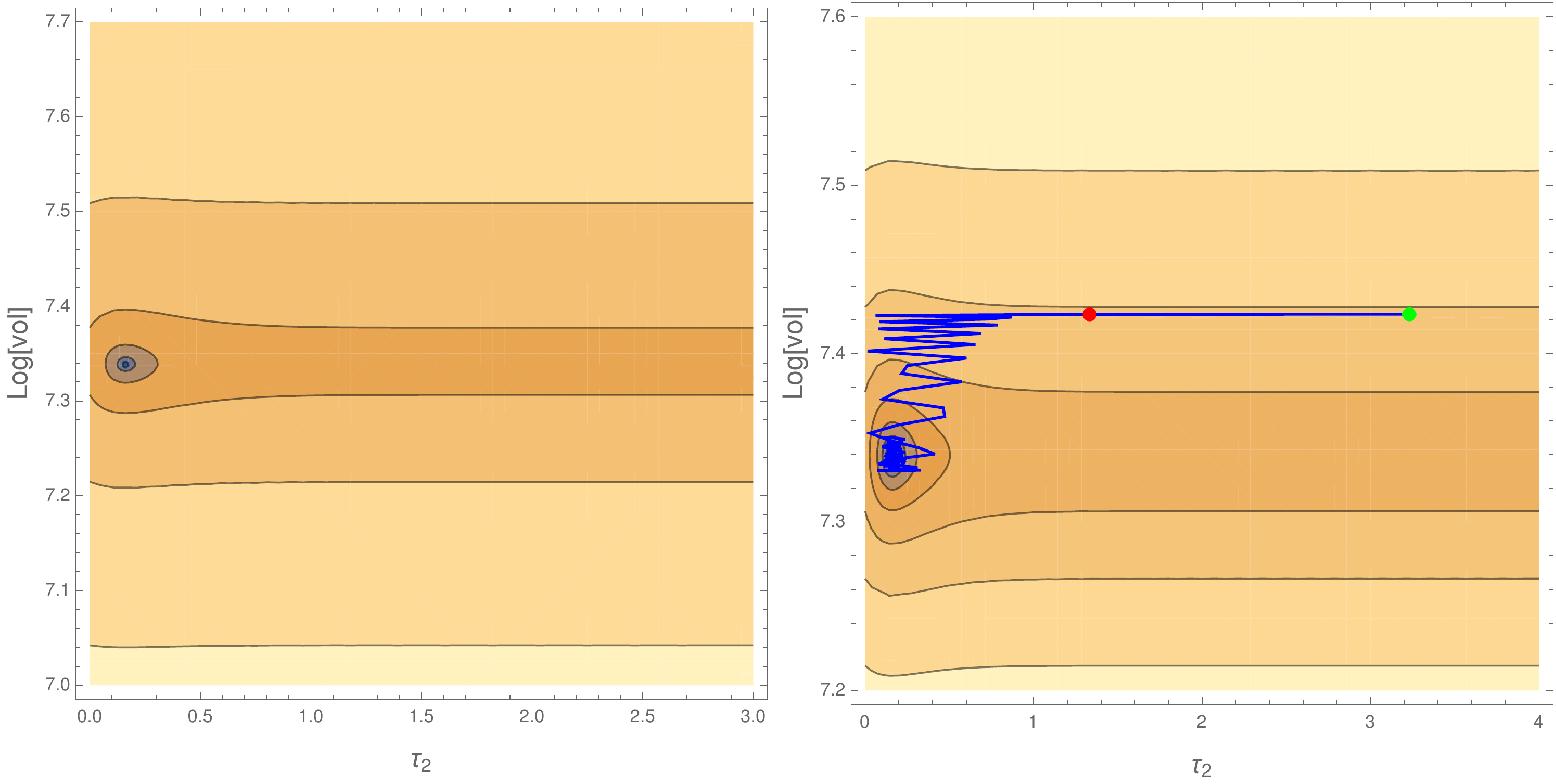}
\caption{Inflationary trajectory in the ($\tau_2$, $\ln\vo$)-plane for the third case in Tab.~\ref{tab5} and \ref{tab6}. The green dot represents horizon exit while the red dot denotes the end of slow-roll inflation where $\epsilon_s=1$.} 
\label{Plots3}
\end{center}
\end{figure}

Let us end this section by focusing on the third case in Tab.~\ref{tab5} and \ref{tab6} which is characterised by rather natural values of the underlying parameters. Fig.~\ref{Plots1} shows the numerical evolution of all fields during the whole inflationary epoch while Fig.~\ref{Plots2} focuses only on the last $8$ efoldings with a clearer representation of the final oscillations around the global Minkowski minimum. Finally Fig.~\ref{Plots3} shows the inflationary trajectory in the ($\tau_2$, $\ln\vo$)-plane. From this plot it is clear that the resulting inflationary dynamics is stable and almost single-field. The distance travelled by the canonically normalised inflaton $\varphi$ during inflation in the single-field approximation is of order $\Delta \varphi\sim M_s \left[\left(\tau_2^{\rm in}\right)^{3/4} - \left(\tau_2^{\rm fin}\right)^{3/4}\right] \sim M_s \sim 0.01\,M_p$, showing that this is a small field model which gives $r\sim 10^{-9}-10^{-8}$. Moreover Fig.~\ref{Plots3} shows that the inflationary energy density shifts the volume away from its global minimum during inflation. This initial misplacement is the origin of the epoch of volume mode domination after the end of inflation \cite{Cicoli:2016olq}. 

\section{Conclusions}
\label{Concl}

Quiver gauge theories play a crucial r\^ole for both the study of gauge/gravity dualities and for phenomenological applications since fractional D3-branes at singularities give rise to chiral matter. However most of the works so far have focused on local constructions in a non-compact background. Whilst these studies can address important model building issues for the visible sector like the realisation of the correct gauge group, chiral spectrum and Yukawa couplings \cite{Aldazabal:2000sa}, they cannot answer global questions regarding moduli stabilisation, supersymmetry breaking and inflation. 

In order to build trustable and fully consistent models, it is therefore essential to embed quiver gauge theories in compact Calabi-Yau three-folds with full moduli stabilisation and an explicit brane set-up and choice of background and gauge fluxes which satisfy tadpole cancellation. We already started this line of research in a series of previous papers \cite{Cicoli:2012vw, Cicoli:2013mpa, Cicoli:2013zha, Cicoli:2013cha} where we embedded in type IIB flux compactifications oriented quiver gauge theories with and without D7 flavour branes. In those models, the original Calabi-Yau three-fold features two identical del Pezzo divisors which are exchanged by the orientifold involution and shrink down to zero size due to D-term stabilisation. Fractional D3-branes then sit at these singularities and can give rise to a visible sector with a trinification, Pati-Salam or MSSM-like gauge group. Closed string moduli stabilisation works as in standard LVS scenarios where an additional del Pezzo divisor supports non-perturbative effects which, together with $\alpha'$ corrections, fix the overall volume exponentially large in string units. This large value of the volume, combined with the fact that in those local models supersymmetry breaking can be sequestered from the visible sector \cite{Blumenhagen:2009gk, Aparicio:2014wxa}, can lead to TeV-scale soft terms. Moreover the cancellation of Freed-Witten anomalies generically imply the presence of non-vanishing gauge fluxes on hidden sector D7-branes which therefore develop a T-brane background that naturally yields a Minkowski (or slightly de Sitter) vacuum \cite{Cicoli:2015ylx}. 

In this paper we extended this analysis by building the first examples of global CY orientifolded quivers. In our setup fractional D3 branes sit at orientifolded singularities in type IIB flux compactifications. These constructions are more generic than the previous ones since they do not require a Calabi-Yau with two identical del Pezzo divisors. Moreover, after Higgsing, local orientifolded quivers can give rise to realistic extensions of the Standard Model without the need of flavour D7-branes \cite{Wijnholt:2007vn}. 

After discussing the general conditions for a consistent global embedding, we presented an explicit Calabi-Yau example where fractional D3-branes live at a dP$_0$ orientifold singularity. This setup can yield an $SU(5)$ GUT-like visible sector whose quantum dynamics generates however a runaway non-perturbative superpotential. This runaway can be avoided by the presence of soft supersymmetry breaking mass terms for the matter fields which would induce a non-vanishing VEV for these modes, resulting in a complete breaking of the visible sector gauge group. Hence we focused on a different D3-brane setup that leads to an $SU(7)\times SO(3)$ (and higher gauge groups) visible sector with no runaway non-perturbative superpotential. The visible sector is chiral and could be broken down to a more realistic gauge group via a proper Higgsing. We leave the construction of more realistic global orientifolded quivers to the future. This paper will serve as a useful reference for the strategy which should be followed to build a consistent global model where a local chiral visible sector is successfully combined with full de Sitter closed string moduli stabilisation in the bulk. 

Besides the dP$_0$ divisor collapsed to a singularity, our explicit Calabi-Yau example features two other dP$_8$ divisors and a large four-cycle controlling the overall volume of the extra dimensions. Both of the rigid del Pezzo's are wrapped by an ED3-instanton while the large divisor supports a hidden T-brane D7-stack that generates a positive contribution responsible for de Sitter uplifting. Closed string moduli stabilisation works again as in standard LVS where non-perturbative effects compete with the leading order $\alpha'$ correction. Supersymmetry is broken by the F-terms of the bulk K\"ahler moduli and it is mediated to the visible sector at the dP$_0$ orientifold singularity via gravitational interactions. 

Interestingly, our model is also able to describe cosmic inflation which can be driven by one of the dP$_8$ moduli. In fact, as soon as this K\"ahler modulus is shifted from its minimum, it immediately features a very flat potential due the exponential suppression of the ED3 contribution \cite{Conlon:2005jm, BlancoPillado:2009nw}. The presence of two blow-up modes is crucial to guarantee the stability of the inflationary dynamics since the volume mode is kept at its minimum during inflation by the dP$_8$ divisor which does not play the r\^ole of the inflaton. In order to fully trust our inflationary model, we performed a complete multi-field numerical evolution following each of the three fields from the initial conditions till the end of inflation where they oscillate and then settle in the Minkowski global minimum. The resulting inflationary model is of small-field type, and so the tensor-to-scalar ratio turns out to be too small to be observed in the near future: $r\sim 10^{-8}$. Moreover, the requirement to reproduce the correct amplitude of the density perturbations fixes the Calabi-Yau volume to a value which is too low to obtain low-energy supersymmetry. Thus in these scenarios supersymmetry does not directly address the hierarchy problem and dark matter cannot be a standard WIMP (either thermally or non-thermally produced) but it arises more naturally from axion-like particles \cite{Cicoli:2012sz}. A promising candidate is the axionic partner of the volume mode.

These models feature also an interesting post-inflationary evolution. The volume mode gets slightly shifted from its global minimum during inflation, and so drives an early period of matter domination after the initial reheating from the inflaton decay. This epoch of modulus domination reduces the number of efoldings to $45$ \cite{Cicoli:2016olq} and leads to a dilution of any previous relic when the volume mode decays \cite{Allahverdi:2013noa, Allahverdi:2014ppa, Aparicio:2015sda, Aparicio:2016qqb, Allahverdi:2016yws}. This late time modulus decay generically causes the production of ultra-light bulk axions which increase the number of effective neutrino-like degrees of freedom, leading to $\Delta N_{\rm eff}\simeq \mc{O}(0.5-1)$ \cite{Cicoli:2012aq, Higaki:2012ar, Cicoli:2015bpq}. This in turn yields a scalar spectral index of order $n_s\simeq 0.97-0.98$, in accordance with Planck data with a non-zero $N_{\rm eff}$ prior \cite{Ade:2015xua}. Let us finally stress that our model represents the first explicit global Calabi-Yau example featuring both an inflationary and a chiral visible sector.

Our work leaves also several open challenges: a systematic classification of viable models, more general del Pezzo singularities with more realistic spectra and couplings, the inclusion of $U(1)$ instantons, a successful model of string inflation with low-scale supersymmetry breaking, etc. At the current level of development of string compactifications, it is important to construct explicit examples which can address concrete physical questions and can also shed some light into potential generalisations and challenges. Our results illustrate how far we have been able to go in exploring realistic string scenarios and they  open up a new avenue to explore physical implications of string compactifications. This is definitely a small but solid step forward.

\section*{Acknowledgments}

We would like to thank Massimo Bianchi, Ralph Blumenhagen, Andres Collinucci, Sven Krippendorf, Raffaele Savelli and Angel Uranga for many useful discussions. PS is grateful to the Bologna INFN division for hospitality when most of this work was carried out. C.M. would like to thank the ICTP Trieste for hospitality. The work  C.M. is supported by the Munich Excellence Cluster for Fundamental Physics `Origin and the Structure of the Universe'. 
The work of M.C. and R.V. is supported by the Programme ``Rita Levi Montalcini for young researchers'' of the Italian Ministry of Research.

\end{document}